\documentclass{aa}

\usepackage{graphicx}
\usepackage{txfonts}
\usepackage{comment}
\usepackage{booktabs}
\usepackage{enumitem}
\usepackage{tikz}
\usepackage{float}
\usetikzlibrary{positioning, arrows.meta}

\usepackage[export]{adjustbox}

\begin{document}

\title{The infrared unmodulated pyramid wavefront sensor: simulations, lab and on sky}

\author{
Muskan Shinde\inst{1}
\and Nicolas Blind\inst{1}
\and Isaac Dinis\inst{1}
\and Sylvain Cetre\inst{2}
\and Angelie Alagao\inst{3}
\and Francisco Oyarzun\inst{3}
\and Romain Fétick\inst{3}
\and Matteo Pasinetti\inst{3}
\and Nathanaël Restori\inst{1}
\and Christophe Lovis\inst{1}
\and Vincent Chambouleyron\inst{3}
\and Benoît Neichel\inst{3}
\and Jean-François Sauvage\inst{3}
\and Thierry Fusco\inst{3}
\and François Huppert\inst{4}
\and Marc Ferrari\inst{4}
}

\institute{
Geneva Observatory, University of Geneva, Chemin Pegasi 51, 1290 Versoix, Switzerland
\and Wakea Consulting, Grenoble, France
\and Aix Marseille Univ, CNRS, CNES, LAM, Marseille, France
\and Aix Marseille Univ, CNRS, Pytheas, OHP, Observatoire de Haute-Provence, France 
}

\date{Received ; accepted }

\abstract
{
Direct imaging of Earth-like exoplanets requires strong suppression of stellar light at small angular separations, achievable with extreme adaptive optics (XAO). This places stringent requirements on wavefront sensors, particularly high sensitivity to low-order aberrations. The pyramid wavefront sensor (PWFS) is widely used for its sensitivity but suffers from nonlinearities and limited dynamic range. Modulation extends the range but reduces sensitivity and limits loop speed. We investigate an unmodulated infrared PWFS (IR-umPWFS), where the dynamic range scales with wavelength, removing the need for modulation and enabling faster operation. To avoid manufacturing issues of the classical 4-sided design, we adopt a 3-sided PWFS, which eliminates the roof defect.
}
{We assess the feasibility and performance of the IR-umPWFS in simulations, lab and on sky, and compare it with a conventional visible modulated PWFS (VIS-mPWFS).}
{The performance of the IR-umPWFS is evaluated through a combination of end-to-end numerical simulations, lab tests, and on-sky observations conducted on the PAPYRUS adaptive optics bench at the Observatoire de Haute-Provence (OHP).}
{
Simulations show reliable operation over a broad range of conditions, reaching Strehl ratios up to $\sim$60\% at 1400\,nm for seeing up to $4''$. Infrared operation improves linearity, reduces optical-gain variations and modal cross-talk. Laboratory results validate the simulations and confirm the pyramid quality. On sky, stable closed-loop operation was achieved over three nights.
Compared with the VIS-mPWFS operating at 500~Hz, the IR-umPWFS
operating at 1000~Hz showed gains of $\sim 1.63$ in Strehl ratio
and $\sim 1.26$ in encircled energy, along with a
$\sim 2.03$-fold reduction in residual wavefront error. This is a system-level comparison, as the two arms differ in several parameters: the pyramid geometry (3- vs 4-sided, the latter with a roof defect), the sensing wavelength and the absence of modulation with the resulting higher loop speed, which are the advantages being tested, and the different non-common path aberrations, whose contribution is removed from the Strehl ratios. The IR-umPWFS remained robust to tip-tilt offsets up to $3\,\lambda/D$ at 1400\,nm.
}
{The IR-umPWFS is a strong candidate for future XAO systems: operating in the infrared extends the linear range sufficiently to remove the need for modulation, enabling higher loop speeds that are no longer limited by the speed of the modulation mirror, which contributed to the significantly improved on-sky performance. The OHP turbulence conditions ($D/r_0 \sim 17$ at 1400\,nm) are comparable to typical Very Large Telescope (VLT) conditions ($D/r_0 \sim 20$), supporting the relevance of these results for VLT/RISTRETTO.}


 \maketitle
\nolinenumbers

\section{Introduction}

Direct imaging and characterization of Earth-like exoplanets is one of the key science goals of current and upcoming ground-based large telescopes. However, such observations require suppression of stellar light by several orders of magnitude at small angular separations.  To image an Earth-like planet orbiting a Sun-like star, a contrast of $10^{-10}$ is required \citep{Snellen2015}. For habitable zone planets around nearby M dwarfs, such contrasts must be achieved at angular separations of only a few $\lambda/D$  \citep{Lovis2017}. This motivates the combination of high-contrast imaging with high-dispersion spectroscopy, into a technique called high-dispersion coronagraphy. In this approach, adaptive optics (AO) and coronagraphy first suppress the stellar light to obtain a raw contrast of $10^{-4}$--$10^{-5}$ at separations of a few $\lambda/D$, then the high spectral resolution allows additional spectral separation of the planetary signal from residual stellar speckles, which is Doppler-shifted relative to the quasi-static stellar speckle field, and enables further suppression of stellar residuals by factors of $10^{-3}$--$10^{-4}$ through cross-correlation techniques \citep{Snellen2015,Wang2017}. Ultimately, performance at small inner working angles is limited by low-order residual wavefront errors (WFE). Extreme adaptive optics (XAO) systems are therefore essential, as they provide the high wavefront quality and stability required to reach high contrast at separations as small as $1$–$2\,\lambda/D$ to fully exploit the angular resolution of current 8--10\,m class telescopes. This will be even more critical for future extremely large telescopes.

RISTRETTO is one such instrument concept that builds upon these capabilities. It is a high-contrast, high-resolution visible spectrograph proposed as a visitor instrument for the Very Large Telescope (VLT) \citep{Chazelas2020,Lovis2024ristretto,Bugatti_2025}. It aims to detect the reflected light from nearby exoplanets in the visible wavelength regime ($620 - 840 \,\mathrm{nm}$) and to characterize their atmospheres. Specifically, it is driven by the goal of characterizing the atmosphere of Proxima b, a rocky exoplanet in the habitable zone of our closest star Proxima Centauri. Proxima b is located at an apparent separation of 37\,mas from its host star, corresponding to $2\,\lambda/D$ at $\lambda = 750\,\mathrm{nm}$ for an 8\,m-class telescope, and has an expected reflected-light contrast on the order of $10^{-7}$ \citep{Lovis2017}. To observe it, a raw contrast of less than $10^{-4}$ at $2\lambda/D$ is required before the spectrograph. An XAO system followed by a Phase-Induced Amplitude Apodizer \& Nuller coronagraph is therefore needed before the spectrograph to meet these requirements. In terms of performance, a Strehl ratio above 70\% at 750\,nm, or a total WFE below 70\,nm RMS \citep{Blind_2025}, and low-order WFE (first 50 Zernike modes) below 10\,nm RMS are required. Meeting these specifications necessitates a wavefront sensor with high sensitivity to low-order aberrations and petal modes.

The pyramid wavefront sensor (PWFS) \citep{Ragazzoni1996} has become the standard choice for XAO systems due to its superior sensitivity compared to Shack–Hartmann sensors \citep{Ragazzoni1999}. It has been adopted in several state-of-the-art instruments, including LBT/FLAO \citep{Esposito2011}, SPHERE/SAXO+ \citep{mazoyer2024}, SCExAO \citep{Martinache2014}, MagAO-X \citep{Males2018,Males2020}, KPIC \citep{Delorme2021}, and GPI~2.0 \citep{limbach2020}. It has, however, a nonlinear response and a limited dynamic range, which is typically increased by using modulation. But this modulation has several disadvantages: it reduces sensitivity to low-order modes \citep{Ragazzoni2002,Agapito2023}, decreases the ability to sense petal-piston modes \citep{BertrouCantou2022,Engler2022}, and makes loop speeds above $\sim$2\,kHz difficult due to the modulation mirror speed limit \citep{Lozi2022}. This motivated us to move away from modulation and toward the study of an unmodulated PWFS (umPWFS). 

The potential of the umPWFS, as well as its limitations, have been studied in previous works. Removing the modulation maximizes the sensitivity of the PWFS, in particular to low-order modes \citep{Ragazzoni1999,Ragazzoni2002,Agapito2023}. In closed loop, the residual wavefront reduces its response through the optical-gain effect \citep{Deo2019}, which can be compensated by optical-gain tracking \citep{Deo2019,Chambouleyron2020,Chambouleyron2021}. These optical-gain variations are less pronounced without modulation, which constitutes an additional advantage of the umPWFS. The main limitation of the umPWFS is its small linear range. Alternatives to modulation have been proposed to extend the linear range, such as static modulation \citep{WangBai2024,Verinaud2024} and nonlinear reconstructors \citep{Frazin2018,LandmanHaffert2020,Wong2023,Landman2024}, but static modulation has only been studied analytically and numerically, and nonlinear reconstruction adds complexity to the control.

\begin{figure}
\centering
\includegraphics[width=0.75\columnwidth]{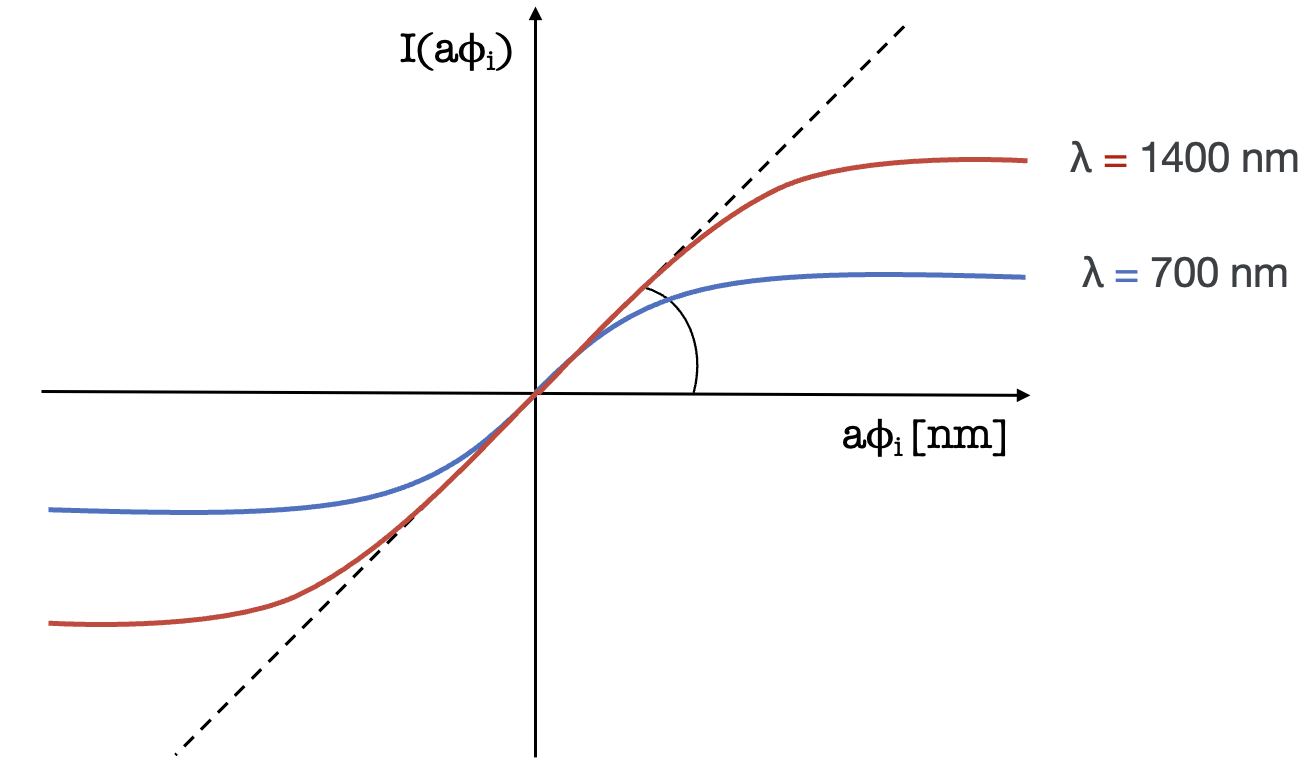}
\caption{Linearity curves of the PWFS at different wavelengths. The x-axis shows the optical path difference amplitude $a$ of a given mode $\phi_i$, while the y-axis shows the corresponding intensity response $I(a\phi_i)$ measured by the PWFS.}
\label{fig:Linearity_curves_with_wavelenght}
\end{figure}

In this work, we adopt a complementary approach that relies on a simple linear reconstruction and control architecture. The key idea is to exploit the wavelength dependence of the PWFS linearity range, which scales linearly with wavelength. Operating the sensor at longer wavelengths naturally extends its linear regime, as illustrated in Fig.~\ref{fig:Linearity_curves_with_wavelenght} \citep{Shinde2024}. This provides a simple and robust way to improve the performance of an unmodulated PWFS, and motivates the study of an infrared umPWFS (IR-umPWFS). 

However, moving the sensing wavelength to the infrared also introduces a trade-off. For a given optical path difference, the corresponding phase aberration decreases with wavelength. As a result, the phase measured by the PWFS is smaller at longer wavelengths, reducing the sensor signal and, therefore, its sensitivity. But several additional factors must also be considered. First, optical gain variations are smaller in the infrared, resulting in a more stable sensor response during closed-loop operation \citep{Shinde2024,Bond2018}. Second, operating at longer wavelengths reduces modal cross-talk, improving the robustness of the wavefront reconstruction \citep{Shinde2024}. Third, the main science targets of RISTRETTO are M dwarfs \citep{Lovis2017}, which are significantly brighter in the infrared than in the visible (e.g., $V = 11.13$ vs. $I = 7.41$ for Proxima; \citealt{Lovis2017}), providing a higher photon flux for wavefront sensing. 

The potential advantages of infrared wavefront sensing have been investigated previously. \citet{Wizinowich2016} compared visible and infrared PWFS and showed that the infrared PWFS achieved lower residual wavefront errors than its visible counterpart. Similarly, \citet{Bond2018} demonstrated through end-to-end simulations that an infrared PWFS can provide improved performance under realistic observing conditions. However, these studies have been largely limited to simulations, and experimental validation under representative observing conditions remains scarce. To our knowledge, this work presents the first on-sky closed-loop operation of an unmodulated 3-sided PWFS in the infrared, together with a side-by-side comparison with a visible modulated PWFS (VIS-mPWFS) on the same telescope. Specifically, we demonstrate (i) the feasibility of the IR-umPWFS in simulations and in the laboratory, (ii) its on-sky operation, and (iii) a comparison with an existing VIS-mPWFS, in which the IR-umPWFS concept (infrared, unmodulated, faster loop speed) is compared as a whole with the standard approach.

In Sect.~\ref{sec:type_of_pyramid}, we present a comparison between the 3-sided and 4-sided PWFSs and justify our choice of pyramid geometry. In Sect.~\ref{sec:simulations_bench}, we present the results obtained from end-to-end simulations as well as laboratory tests performed to evaluate the performance of the IR-umPWFS. In Sect.~\ref{sec:on_sky}, we present the results from on-sky tests conducted on the PAPYRUS (Provence Adaptive oPtics PYramid RUn System) bench at the Observatoire de Haute-Provence (OHP), assessing the sensor performance under real observing conditions and its comparison with the VIS-mPWFS. Finally, in Sect.~\ref{sec:conclusions}, we summarize the main results and present the conclusions of this work.

\section{Type of pyramid}
\label{sec:type_of_pyramid}

\begin{figure}
\centering
\includegraphics[width=0.5\columnwidth]{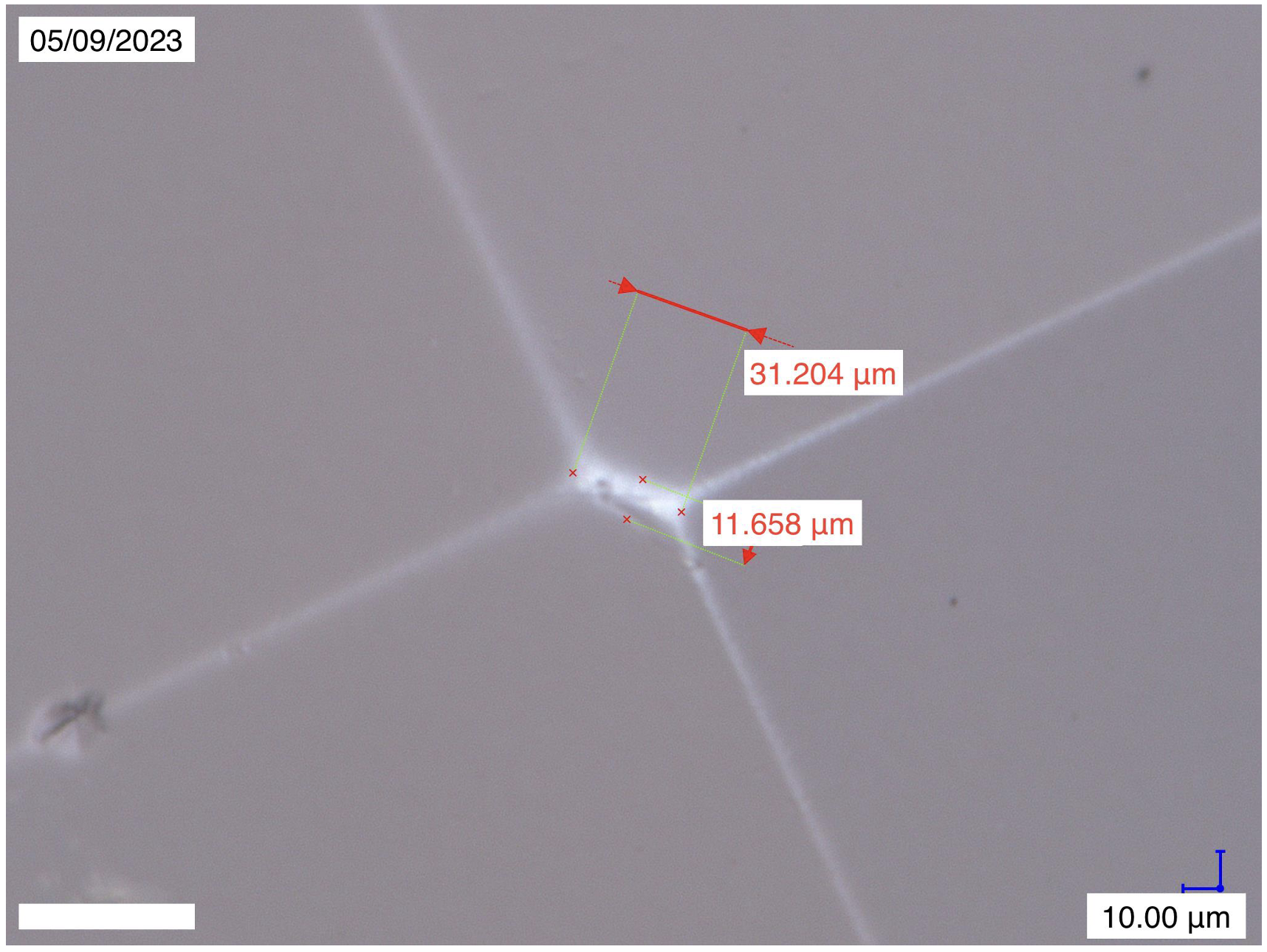}
\caption{Microscope image of a 4-sided pyramid optic. The thin bright lines correspond to reflections from the pyramid edges. At the center, instead of forming a sharp apex, the pyramid exhibits a roof defect caused by unequal polishing depths of the pyramid faces. The measured dimensions of this defect are approximately 31.2\,$\mu$m along the roof ridge and 11.7\,$\mu$m across the apex region. Image courtesy: PAPYRUS team/LAM.}
\label{fig:4pwfs}
\end{figure}

In the umPWFS, the point spread function (PSF) is focused directly onto the pyramid apex; therefore, the quality of the pyramid tip is a critical parameter. The classical pyramid used in this type of sensor is a 4-sided pyramid, in which unequal polishing depths of the pyramid faces often produce a roof defect rather than a sharp apex, as shown in Fig.~\ref{fig:4pwfs}. The roof defect should be less than $0.25\,\lambda_{\mathrm{WFS}}/D$ to obtain balanced pupil images with this type of pyramid, where $\lambda_{\mathrm{WFS}}$ is the wavelength of the wavefront sensor (WFS). Achieving such accuracy is challenging, even with significant manufacturing effort from the companies we contacted. In the pyramid shown in Fig.~\ref{fig:4pwfs}, the roof defect extends over approximately $31\,\mu\mathrm{m}$ along the ridge. To operate with such a pyramid, the PSF size at the apex would need to be about $120\,\mu\mathrm{m}$, implying a focal ratio greater than $f/200$ at a wavelength of $639\,\mathrm{nm}$, which is hard to work with.

\begin{figure}
\centering
\includegraphics[width=1\columnwidth]{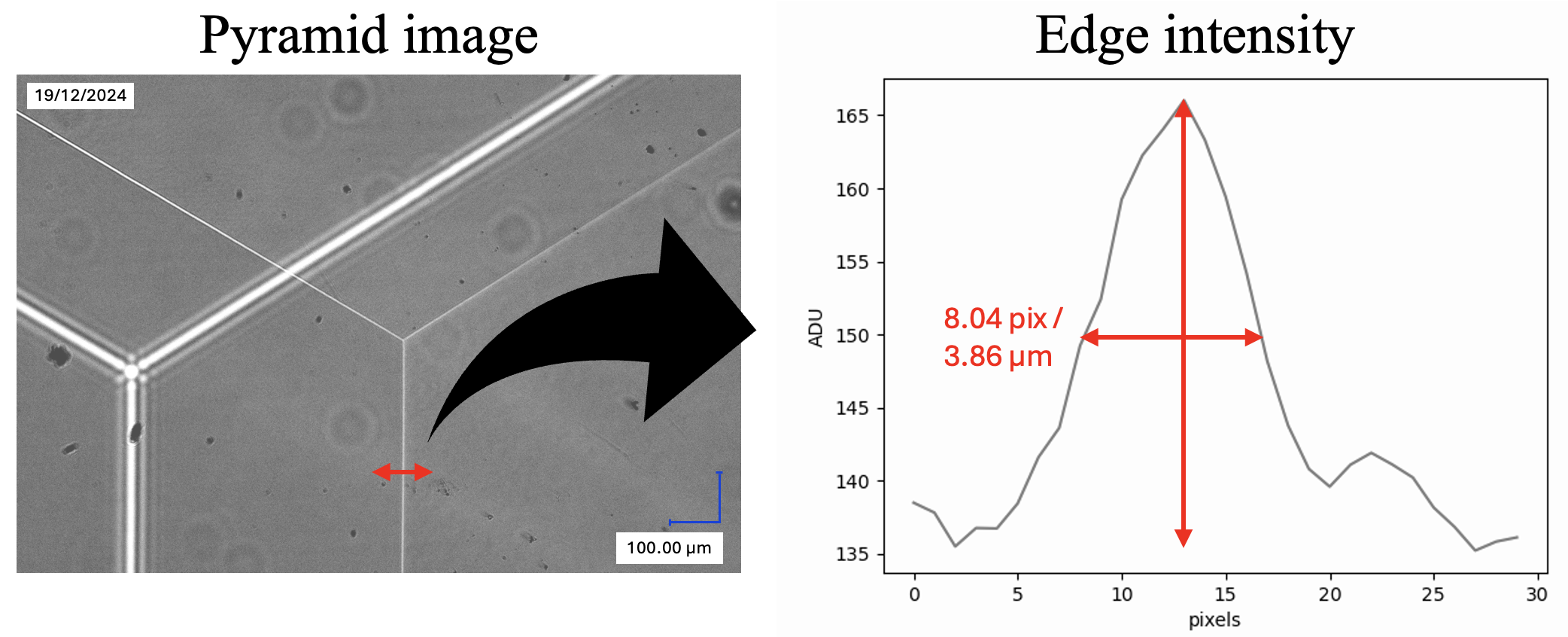}
\caption{Left: Microscope image of the apex region of the 3-sided pyramid optic. The bright thick bands visible on the left correspond to back-reflections from the pyramid faces, while the thin bright lines at the center mark the physical pyramid edges. A short vertical intensity cut across one of the edges (red double arrow) is extracted to evaluate the optical quality of the pyramid tip. Right: Intensity profile measured along the cut shown in the left panel. The width of the pyramid edge is estimated from the full width at half maximum (FWHM) of the intensity peak, yielding a value of 8.04 pixels (3.86\,$\mu$m).}
\label{fig:3pwfs}
\end{figure}

As an alternative, a 3-sided pyramid wavefront sensor can mitigate this issue, as it does not suffer from the same roof defect (Fig.~\ref{fig:3pwfs}). The pyramid shown is a glass prism manufactured by AllInOptics, with a small apex angle of less than $2^\circ$. No roof defect is present, and the measured edge (and apex) width is better than $3.86\,\mu$m (measurement is limited by the diffraction of the microscope used for the inspection). This level of accuracy allows operation at focal ratios as low as $f/24$. 

We performed simulations to test the performance, under the same conditions, of 3-sided and 4-sided PWFSs on several aspects including sensitivity, dynamic range and optical gains, and they showed very similar performance \citep{Shinde2024}. This motivated us to use a 3-sided PWFS for our studies without impact on performance; for the same reason, the pyramid geometry is not expected to bias the on-sky comparison presented in Sect.~\ref{sec:on_sky}.

\section{Simulations and lab bench}
\label{sec:simulations_bench}

\begin{figure*}
\centering

\includegraphics[ width=0.8\textwidth]{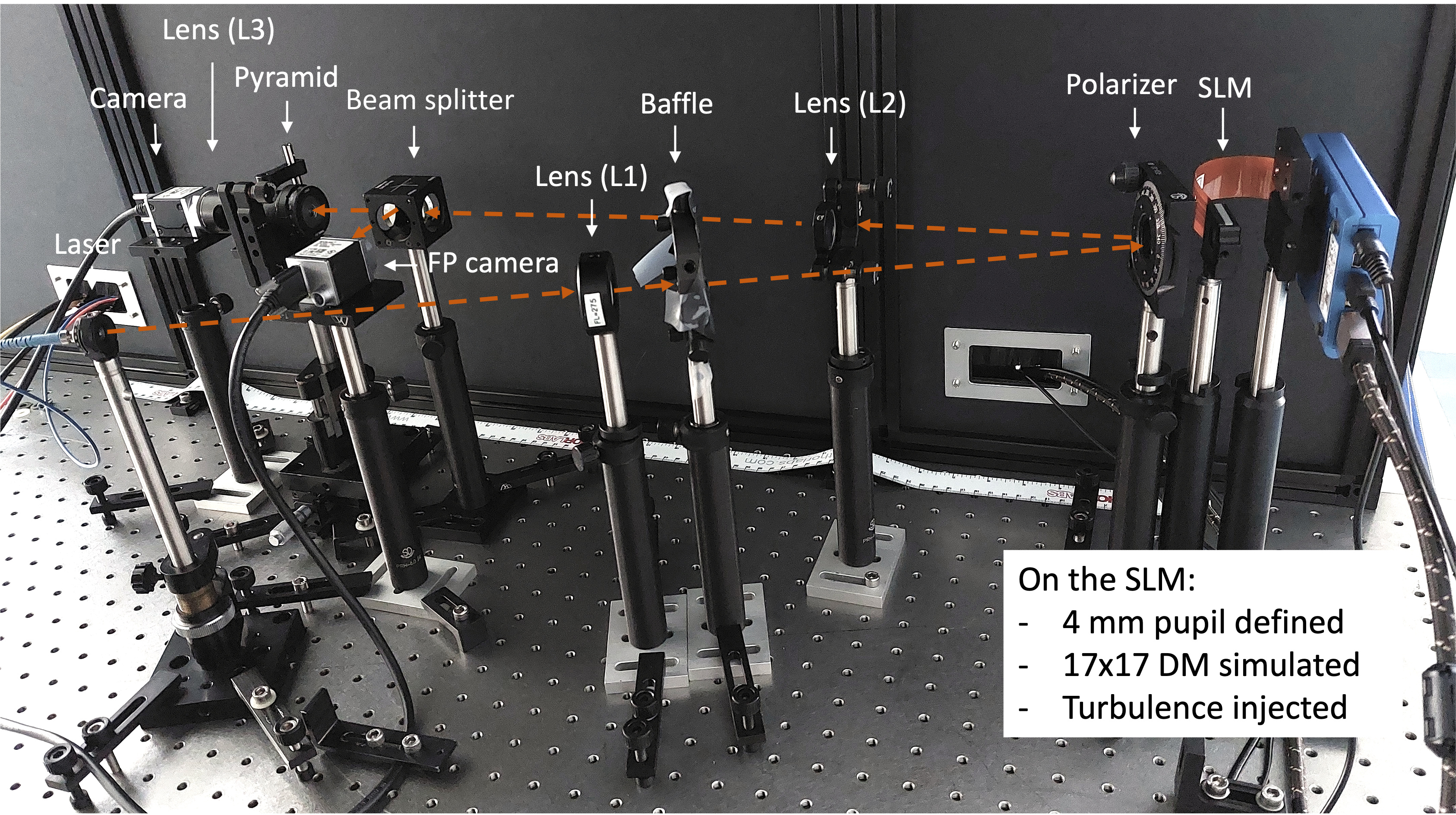}

\caption{Layout of the adaptive optics bench used to test the IR-umPWFS. The light source is a diode laser operating at 639\,nm. The beam is collimated by Lens L1, passes through a baffle to suppress stray light, and through a linear polarizer to ensure proper polarization for the reflective Holoeye spatial light modulator (SLM). The SLM introduces phase aberrations and performs wavefront correction by acting as a deformable mirror. The beam is then focused by Lens L2 onto the tip of a refractive 3-sided pyramidal prism (see Fig.~\ref{fig:3pwfs}). A beam splitter placed near the focal plane directs a fraction of the light to a focal-plane camera for monitoring, while the remaining light falls on the pyramid tip. The resulting pupil images are re-imaged onto the PWFS camera using Lens L3.}
\label{fig:AO_bench}
\end{figure*}

Here we describe the end-to-end numerical simulations and the AO lab bench used to evaluate the performance of the IR-umPWFS.

\begin{table}[t]
\centering
\caption{Summary of the simulation parameters}
\label{tab:simulation_parameters}
\small
\setlength{\tabcolsep}{4pt}
\resizebox{\columnwidth}{!}{%
\begin{tabular}{llc}
\toprule
\textbf{Category} & \textbf{Parameter} & \textbf{Value} \\
\midrule

Telescope
& pupil diameter [m] & 1.52 \\
& pupil sampling [pix] & 500 \\
& pupil grid sampling [pix] & 550 \\
& obscuration ratio [\%] & 25 \\

\midrule
Turbulence
& profile & Kolmogorov + Taylor frozen-flow \\
& outer scale [m] & 40 \\

\midrule
DM
& no. actuators & $17\times17$ \\
& valid actuators & 240 \\
& modal basis & tip-tilt + 183 KL \\

\midrule
PWFS
& no. sub-apertures & $40\times40$ \\
& sub-pupil separation [pix] & 60 \\

\midrule
Control
& approach & modal \\
& reconstruction matrix & truncated pseudo-inverse \\
& temporal control & leaky integrator (1\% leak)\\
& frame rate [Hz] & 1000 \\
& delay [frames] & 0--2 \\

\bottomrule
\end{tabular}
}
\end{table}

\subsection{Simulation description}
\label{sec:simulation_description}

The AO simulations are implemented in Python, with the wavefront sensor, deformable mirror (DM), and all optical elements defined on a common pupil grid. Optical propagation and field representations are done using the \texttt{HCIPy} framework \citep{Por2018HCIPy}, while PWFS specific components are implemented through a dedicated custom package. The telescope entrance pupil is generated and sampled on a square grid ($n_{\mathrm{pupil}} = 550$, corresponding to the on-bench configuration in Sect.~\ref{sec:AO_bench_description}). The DM is modeled as a set of Gaussian influence functions arranged on a $17 \times 17$ square actuator grid, corresponding to the on-bench configuration in Sect.~\ref{sec:AO_bench_description}, with a specified inter-actuator coupling of $30\%$ which is representative of typical ALPAO DMs. The DM is used both to introduce phase aberrations and to apply corrective commands.  Wavefront sensing is performed using a 3-sided PWFS model, forward propagation through which produces re-imaged pupils on a detector grid ($n_{\mathrm{PWFS}} = 40$ pixels per pupil diameter), separated by 1.5 pupil diameters to prevent overlap. The simulations are monochromatic: the WFS operates at $\lambda_{\mathrm{WFS}} = 1400$\,nm, and the PSF, from which the Strehl ratio is computed, is evaluated at the same wavelength, $\lambda_{\mathrm{PSF}} = 1400$\,nm.

The PWFS signal processing and sensor calibration follow the approach of meta-intensities described in \citet{Chambouleyron2023}. For an input phase $\phi$, the raw PWFS image $I(\phi)$ is first multiplied by a valid-pixel mask $M$ to exclude detector regions that do not contain useful pupil information. The masked image is then normalized by its total flux to obtain the normalized intensity $\tilde{I}(\phi)$: 
\begin{equation}
\tilde{I}(\phi) = \frac{I(\phi)\, M}{\sum\limits_{\mathrm{pixels}} I(\phi)\, M}
\end{equation}

\noindent The same procedure is applied to the reference image $I_0$, corresponding to a flat wavefront, yielding the normalized reference intensity $\tilde{I}_0$. The slope is then defined as the difference between these normalized intensities:
\begin{equation}
S(\phi) = \tilde{I}(\phi) - \tilde{I}_0 .
\end{equation}

 Calibration is performed using a push-pull interaction matrix procedure based on an orthogonal Karhunen-Lo\`eve (KL) modal basis, including tip and tilt modes. For each KL mode $\phi_i$, we used a value of calibration amplitude $a = 0.05\, \lambda_{\mathrm{WFS}}$. To obtain the push-pull response, slopes are first computed from the PWFS images, and their difference is then taken:

\begin{equation}
\delta S(\phi_i) = \frac{S(a\,\phi_i) - S(-a\,\phi_i)}{2a}
\end{equation}

\noindent The resulting responses are put into the interaction matrix,
\begin{equation}
\mathcal{D} = \left[ \delta S(\phi_1), \dots, \delta S(\phi_N) \right].
\end{equation}

\noindent The reconstruction matrix is obtained as a truncated pseudo-inverse of $\mathcal{D}$ using a threshold parameter,
\begin{equation}
\mathcal{R} = \mathcal{D}^{\dagger},
\end{equation}

\begin{figure}
\centering
\includegraphics[width=0.24\textwidth]{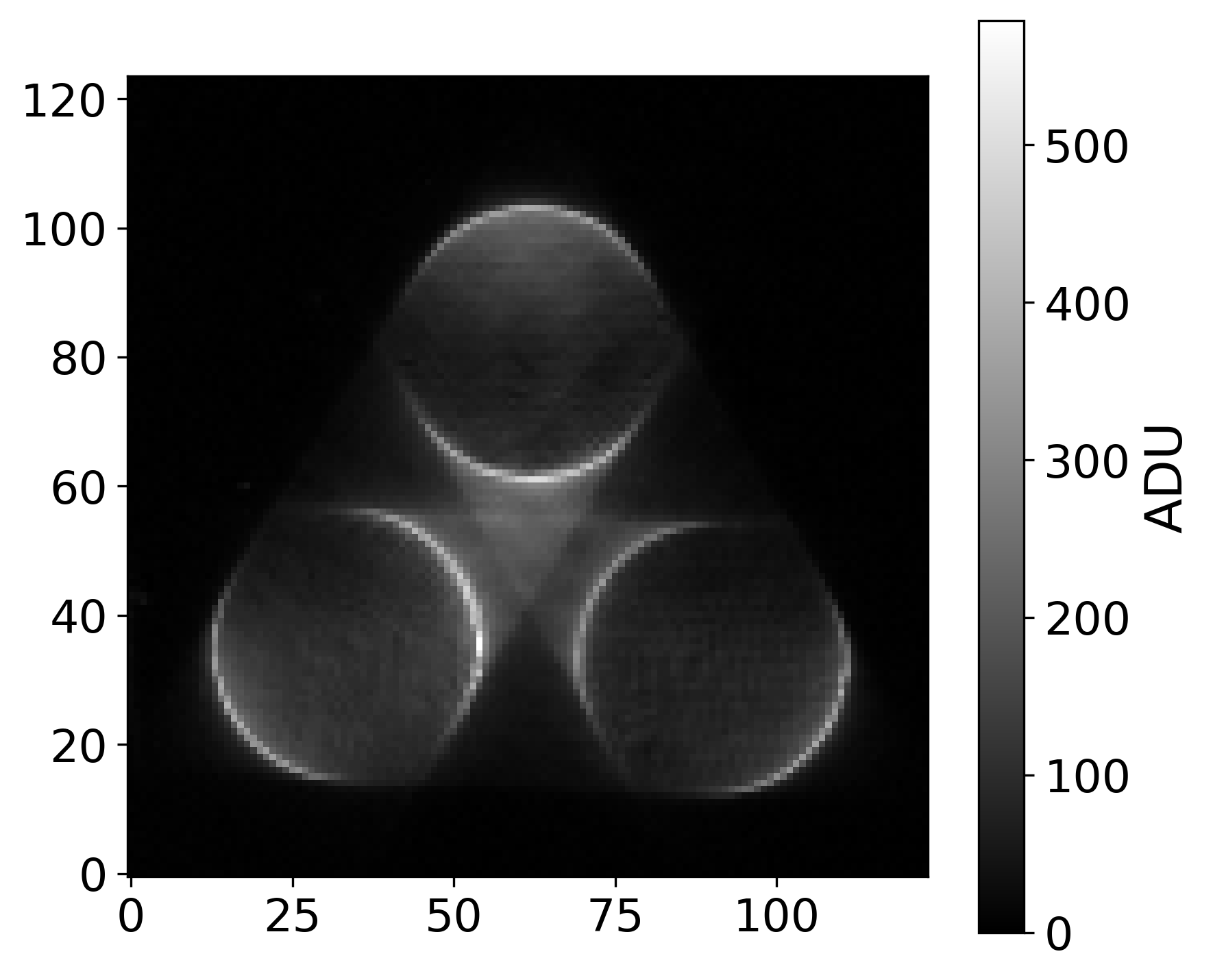}
\centering
\includegraphics[width=0.24\textwidth]{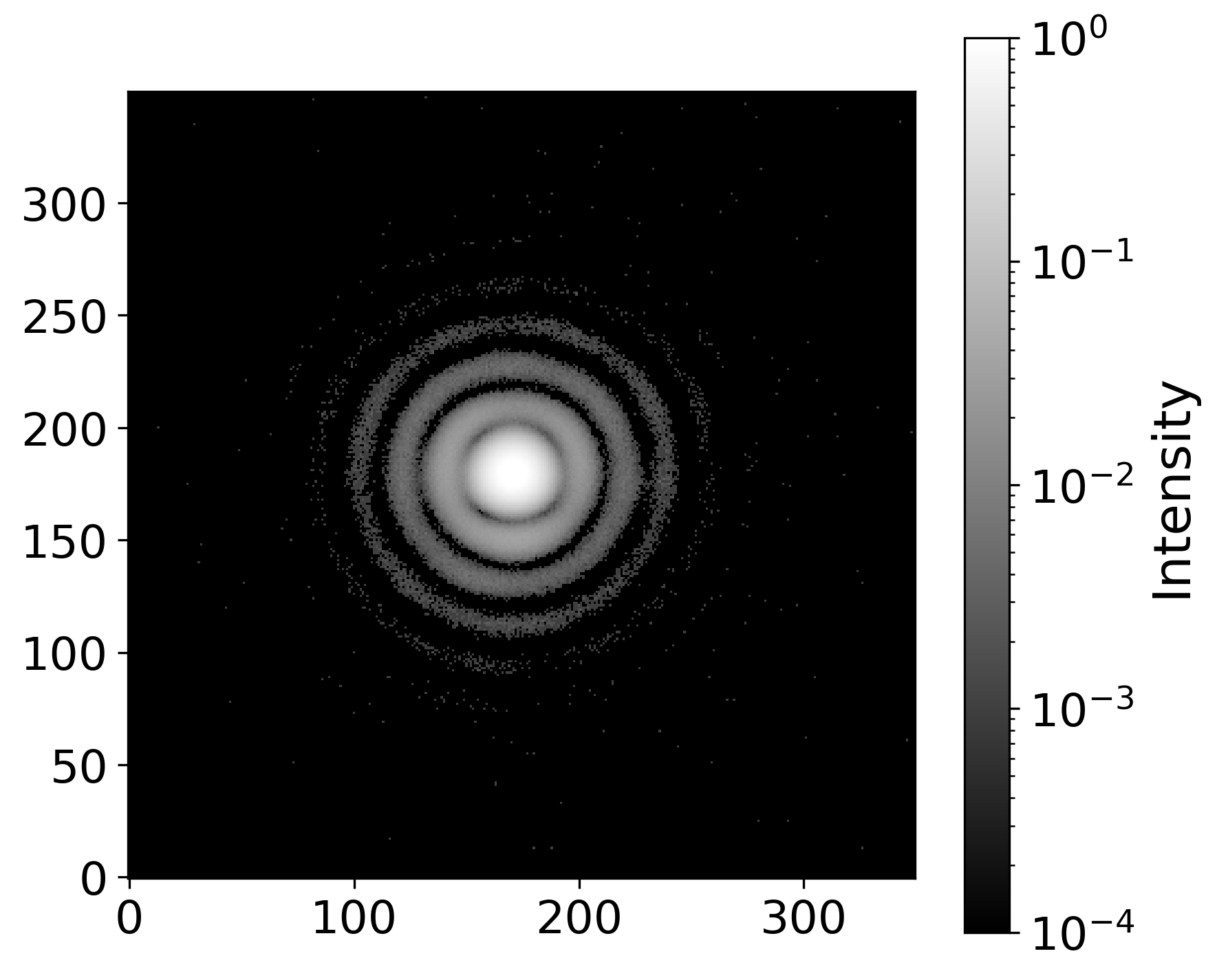}

\caption{Left: The 3-sided PWFS reference image after detector binning, displaying the three pupil images, each sampled with $\sim$42 pixels across the pupil diameter. Right: Corresponding PSF obtained on the AO bench for a flat wavefront, showing a bright central core with visible Airy rings.}
\label{fig:3PWFS_combined}

\end{figure}

\begin{figure*}
\centering
\includegraphics[width=\textwidth]{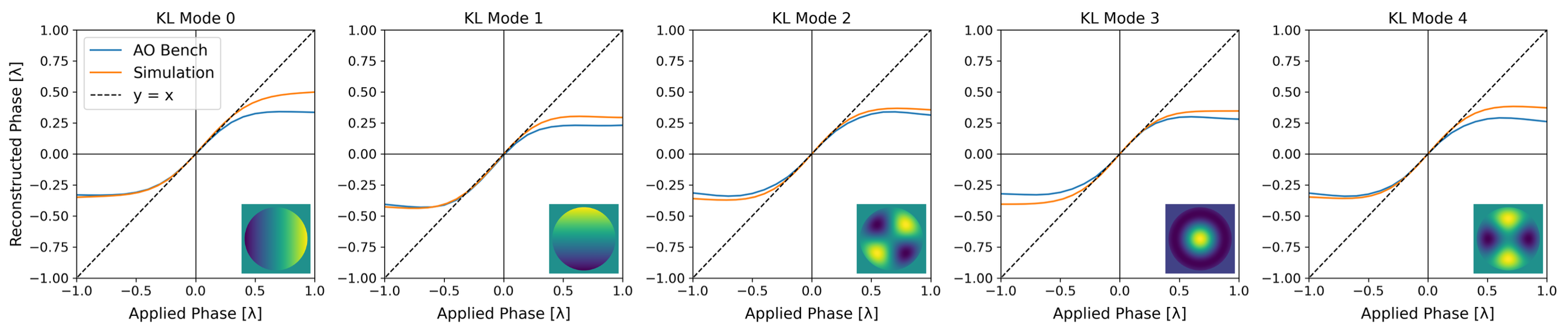}

\caption{Linearity curves of the first five KL modes, comparing end-to-end simulations at $\lambda_{\mathrm{WFS}} = 1400$\,nm (orange) with measurements obtained on the AO bench at $\lambda_{\mathrm{WFS}} = 639$\,nm (blue). Both are subjected to the same phase in radians (see Sect.~\ref{sec:AO_bench_description}). The input and reconstructed phase amplitudes are expressed in units of $\lambda_{\mathrm{WFS}}$, where $1~\lambda_{\mathrm{WFS}}$ corresponds to $2\pi$~rad of phase. The dashed line indicates the ideal linear response ($y=x$). The first two modes correspond to tip-tilt. The corresponding KL mode phase is shown in the lower-right corner of each plot.}

\label{fig:linearity_curves_bench_vs_simulations}
\end{figure*}

\noindent This provides a linear reconstructor mapping PWFS measurements onto modal coefficients. To improve the stability of the reconstruction, we computed the pseudo-inverse using a threshold of $10\%$ of the maximum singular value. Singular values below this threshold were discarded to avoid amplifying noise associated with poorly sensed eigenmodes and preventing unstable commands on the DM. The threshold value of \(10\%\) was selected because it yielded the best performance among the tested values.

Closed-loop control is implemented as a standard integrator with configurable loop gain $g$, leakage factor $\ell$, and frame delay $d$. At a given iteration $k$, the current residual phase is propagated through the PWFS optics to generate the PWFS signal. The slope $S$ is projected onto modal coefficients using the reconstruction matrix, and the DM command is updated according to:
\begin{equation}
\mathbf{a}_{k+1} = (1 - \ell)\,\mathbf{a}_k + g\,\mathcal{R}\,S_{k-d}
\end{equation}

\noindent Typical values used were $g = 0.4$--$1.0$ for the loop gain, $d = 0$--$2$ frames for the delay, and a leakage factor of $\ell = 0.01$. The simulation parameters are described in Table~\ref{tab:simulation_parameters}.

\subsection{AO bench description}
\label{sec:AO_bench_description}

\begin{table}[t]
\centering
\caption{Description of the AO bench components}
\label{tab:ao_bench_description}
\small
\setlength{\tabcolsep}{4pt}

\resizebox{\columnwidth}{!}{%
\begin{tabular}{llc}
\toprule
\textbf{Category} & \textbf{Parameter} & \textbf{Value} \\
\midrule

Light source
& model & diode laser \\
& wavelength [nm] & 639 \\

\midrule
Pupil
& implementation & simulated on SLM \\
& diameter [mm] & 4 \\

\midrule
SLM
& model & Holoeye ERIS, $1920\times1200$ \\
& & 8\,$\mu$m pitch, 60\,Hz \\

\midrule
Cameras
& science camera & Basler a2A1920-160umBAS \\
& PWFS camera & Basler a2A2840-48umBAS \\

\midrule
Optics
& Lens 1 focal length [mm] & 275 \\
& Lens 2 focal length [mm] & 400 \\
& Lens 3 focal length [mm] & 30 \\
& beam splitter & 50/50 cube \\

\midrule
DM
& implementation & simulated on SLM \\
& no. actuators & $17\times17$ \\
& valid actuators & 240 \\
& modal basis & tip-tilt + 183 KL \\

\midrule
PWFS
& optic & 3-sided glass pyramidal prism \\
& no. sub-apertures & $42\times42$ \\
& sub-pupil separation [pix] & 56 \\

\bottomrule
\end{tabular}%
}
\end{table}

The AO bench to test the IR-umPWFS is shown in Fig.~\ref{fig:AO_bench}, and the components of the bench are described in Table~\ref{tab:ao_bench_description}. The pyramid element is a refractive 3-sided glass prism as described in Sect.~\ref{sec:type_of_pyramid}. The light source is a diode laser operating at a wavelength of 639\,nm. After passing through Lens 1, the collimated laser beam passes through a baffle to cut the excess light before illuminating the Holoeye spatial light modulator (SLM), which defines a 4\,mm circular pupil. To define a pupil on the SLM, a phase tilt is applied within the pupil region to shift the diffracted beam off-axis, and separate it from the zero-order diffraction of the SLM. Outside the pupil, a strong tilt is applied to deflect the light further away from the optical axis, effectively removing unwanted light from the region of interest. The SLM has three roles within the system: (i) pupil definition, (ii) introduction of phase aberrations/turbulence, and (iii) acting as a DM. A linear polarizer in front of the SLM is used in double-pass. The reflected beam is focused at $f/100$ over the pyramid apex, resulting in closely spaced pupil images. Pupils are sampled with 128 pixels. In practice, pixels are binned by 3, resulting in a sampling of 42 pixels. The PWFS image obtained for a flat wavefront is shown in Fig.~\ref{fig:3PWFS_combined} (left), where the intensity distribution among the three pupils appears quite uniform, indicating excellent optical quality. A 50/50 beam splitter is positioned just before the focal plane to also reflect light toward a focal plane camera (Fig.~\ref{fig:3PWFS_combined} (right)), which also shows the excellent optical quality of the bench.

Since the same monochromatic laser feeds both the PWFS and the focal-plane camera, $\lambda_{\mathrm{WFS}} = \lambda_{\mathrm{PSF}} = 639$\,nm on the bench, whereas the simulations are performed at $\lambda_{\mathrm{WFS}} = \lambda_{\mathrm{PSF}} = 1400$\,nm. The response of a PWFS and the Strehl ratio depend only on the phase in radians. We therefore emulate operation at 1400\,nm by applying on the SLM the phase screens computed at 1400\,nm, expressed in radians, so that the bench sees exactly the same phase as the simulations. All bench results are thus expressed in units of the simulation wavelength ($\lambda_{\mathrm{WFS}} = \lambda_{\mathrm{PSF}} = 1400$\,nm), i.e., as 1400\,nm-equivalent phases and Strehl ratios.

Real-time control is provided by the Durham Adaptive Optics (DAO) real-time controller (RTC) \citep{barr_2025_17264152}, running on Linux and using shared-memory interfaces to ensure low-latency communication between hardware and software components. This choice was motivated by our intention to perform tests on the PAPYRUS bench which was already equipped with the DAO RTC. In addition, the DAO RTC is also planned to be used for the RISTRETTO instrument.

The PWFS signal processing and calibration on the bench follow the same procedure and parameter values as in the simulations (Sect.~\ref{sec:simulation_description}), to enable a direct one-to-one comparison. The only additional step is a centering routine required to accurately position the PSF on the pyramid tip. Once centered, the bench exhibits excellent stability, remaining well centered for up to one week if left undisturbed.

\subsection{Performance comparison}

Performance tests were performed under PAPYRUS observing conditions (1.52\,m telescope, seeing of 1--4$''$, wind speeds of 2--6\,m\,s$^{-1}$), adopting the same set of parameters in both the end-to-end simulations and the AO bench tests. The corresponding parameters are summarized in Table~\ref{tab:simulation_parameters}.

\subsubsection{Linearity curves}
\label{sec:linearity_curves_sim_bench}

We compared the linearity curves of the first five KL modes obtained from simulations and from the AO bench, as shown in Fig.~\ref{fig:linearity_curves_bench_vs_simulations}. For all modes, the linearity behavior follows a similar trend in both cases, with a very slightly reduced linearity range measured on the bench. This difference can be attributed to the use of a perfect pyramid in the simulations compared with the real one in the lab. Despite these differences, the overall agreement between the simulated and lab linearity curves is very good, indicating the good optical quality of the pyramid element used in the lab.

The linearity curves of the KL mode 1 (tip-tilt) exhibit a noticeable asymmetry around zero. This behavior is strongly dependent on the relative orientation between the 3-sided pyramid and the tip-tilt mode direction. A similar effect is present for other KL modes that are not symmetric in both the x and y directions; however, it is less pronounced than in the tip-tilt case. The best case configurations are when one edge of the 3-sided pyramid is oriented along one of the tip-tilt mode directions.

\begin{figure}
\centering
\includegraphics[width=1\columnwidth]{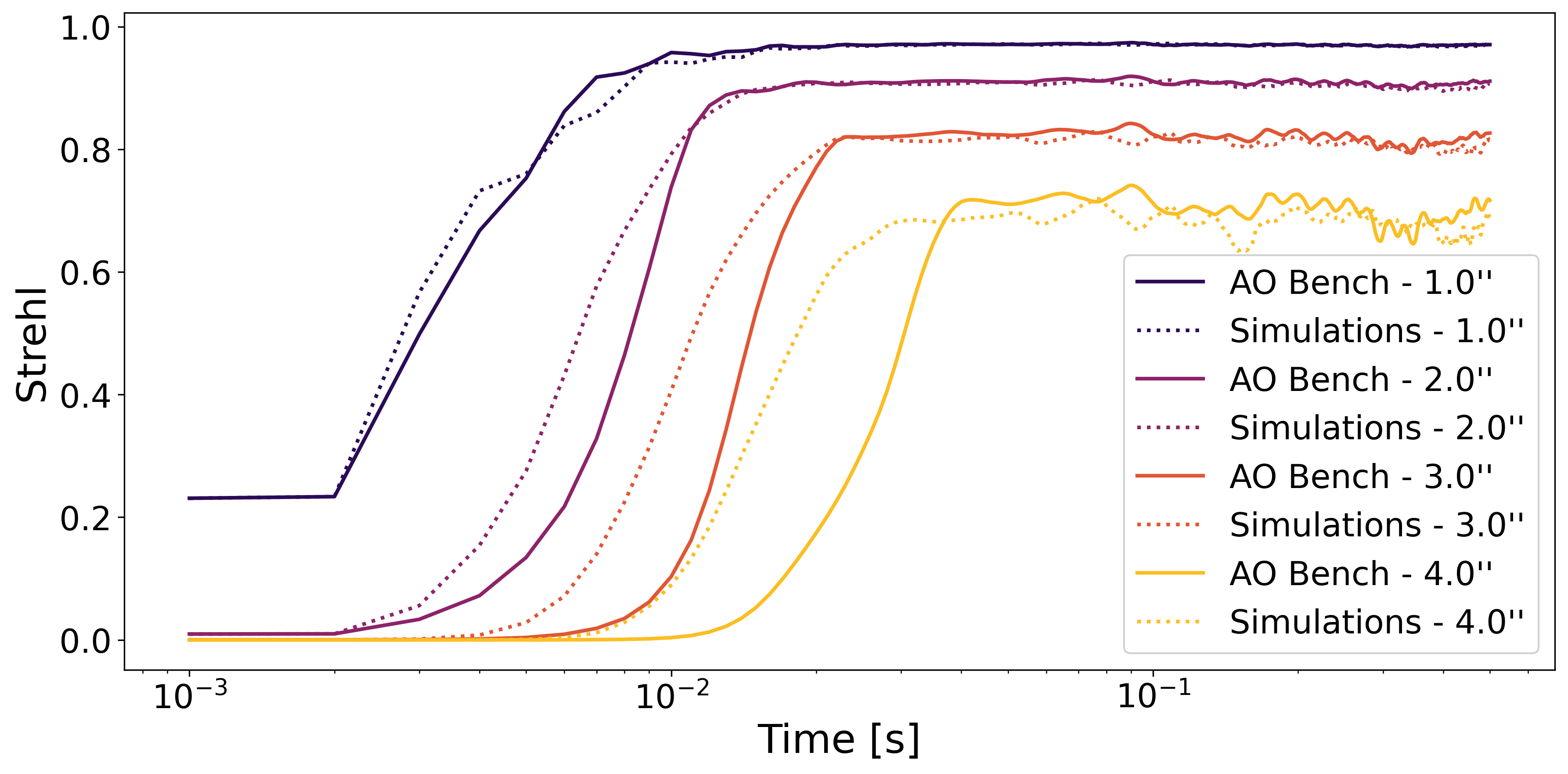}
\caption{Comparison between end-to-end simulations and AO bench measurements under PAPYRUS observing conditions. Strehl ratio time series at $\lambda_{\mathrm{PSF}} = 1400\,\mathrm{nm}$ (on the bench, operation at 1400\,nm is emulated with the 639\,nm laser by applying on the SLM the phase screens computed at 1400\,nm, see Sect.~\ref{sec:AO_bench_description}) for seeing values from 1 to 4\,arcsec, with a one-frame delay and a loop gain of 0.8.}
\label{fig:strehl_timeseries}
\end{figure}

\begin{figure*}
\centering
\includegraphics[width=0.7\textwidth]{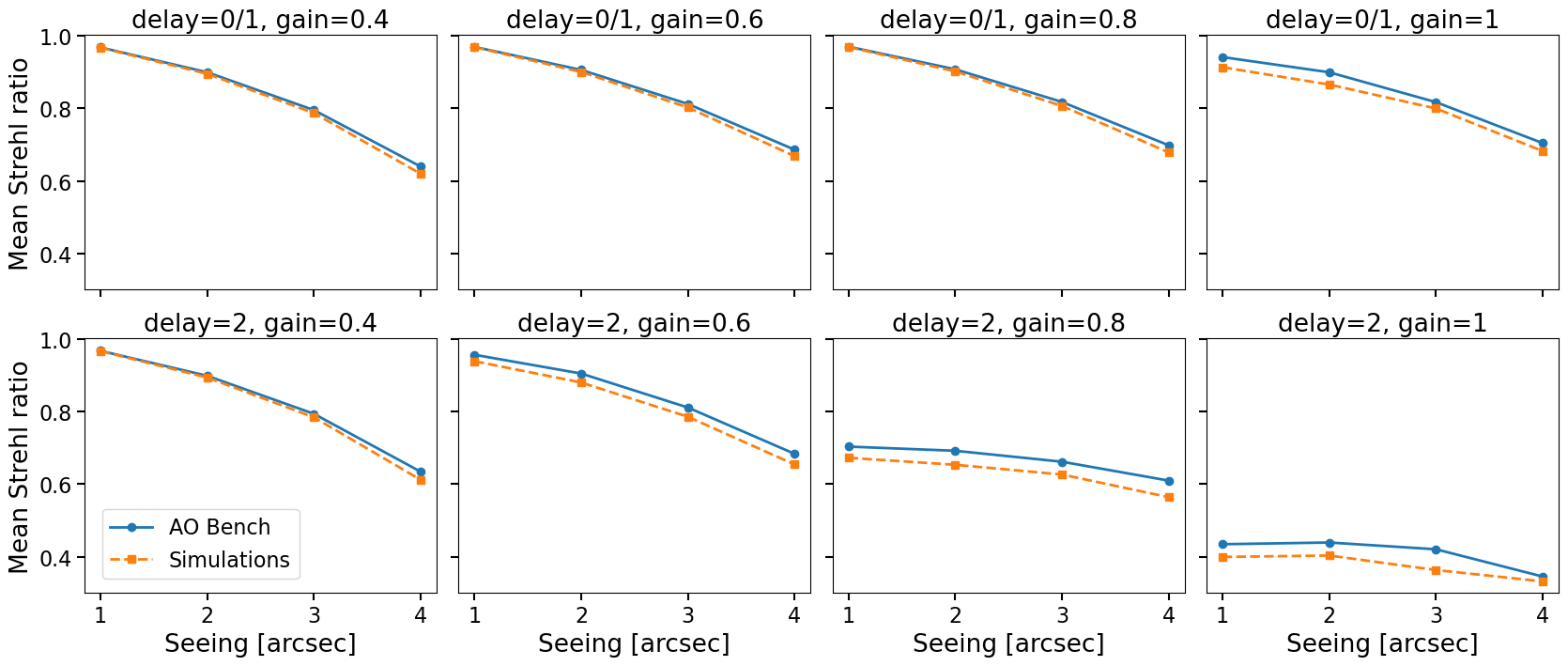}
\caption{Mean Strehl ratio as a function of seeing for different loop delays and gains, comparing end-to-end simulations with AO bench measurements under PAPYRUS observing conditions at $\lambda_{\mathrm{PSF}} = 1400\,\mathrm{nm}$ (on the bench, operation at 1400\,nm is emulated with the 639\,nm laser by applying on the SLM the phase screens computed at 1400\,nm, see Sect.~\ref{sec:AO_bench_description}).}
\label{fig:strehl_vs_seeing}
\end{figure*}

\subsubsection{Closed-loop tests}
\label{sec:closed_loop_tests}
We conducted both end-to-end simulation and lab closed-loop tests under PAPYRUS observing conditions with AO parameters as described in Table~\ref{tab:simulation_parameters}.

The IR-umPWFS successfully closed the AO loop under varying turbulence conditions, with a coherence time of 5\,ms and wind speeds ranging from 2 to 6\,m\,s$^{-1}$. Under seeing conditions up to $4''$, the system achieved Strehl ratios of approximately 60\% at $\lambda_{\mathrm{PSF}} = 1400\,\mathrm{nm}$, which is lower than the ideal Strehl of $\sim$85\% predicted from the residual phase variance derived from the Noll formula considering just fitting error \citep{Noll1976}. The lab results are in good agreement with the end-to-end simulations, as shown in Fig.~\ref{fig:strehl_timeseries}, which presents the temporal evolution of the Strehl ratio. The example corresponds to a configuration with a one-frame delay and a loop gain of 0.8. The effect of the delay is clearly visible, with the correction applied after 1\,ms. The loop converges in less than $\sim$10\,ms for good seeing conditions (1$''$ and 2$''$), while convergence is slower for poorer seeing (4$''$). After convergence, the loop remains stable, and the measured performance closely matches the simulations.

For a seeing of $4''$ at the OHP site, the PAPYRUS configuration corresponds to $D/r_0 \sim 17$ at 1400\,nm. This value is comparable to typical VLT conditions of $D/r_0 \sim 20$ at 1400\,nm, demonstrating that the performance measured on the PAPYRUS bench is representative of what can be expected at the VLT.

To investigate the impact of temporal delay on system performance, we tested delays of 0, 1, and 2 frames across a range of loop gains. The results are summarized in Fig.~\ref{fig:strehl_vs_seeing}. As expected, the Strehl ratio decreases with increasing delay. We note that the performance for delays of 0 and 1 frame is very similar across all gains and seeing conditions; these cases are therefore combined and shown together in the same panels. Optimal performance is obtained for the combinations (delay 0, gain 1), (delay 1, gain 0.8), and (delay 2, gain 0.6). Interestingly, for a delay of 1 frame, the loop remains stable even at a gain of 1, whereas from theory the stability should be limited to gains below 0.8. A similar behavior is observed for a delay of 2 frames, where gains above 0.6 remain stable. This indicates the presence of optical gains in the system. Since the system operates in the infrared, where optical gains are relatively high and have flat variation across spatial frequencies (see Appendix~\ref{sec:optical_gains_appendix}), a mode per mode gain compensation is not required \citep{Shinde2024}. Instead, the optical gain can be effectively incorporated as a global scaling factor within the loop gain which we observe here.

\section{On-sky tests at OHP}
\label{sec:on_sky}

To validate the performance of the IR-umPWFS under realistic observing conditions, on-sky tests were conducted using the PAPYRUS adaptive optics bench at the OHP. We describe here the PAPYRUS bench, the integration of the new infrared arm into the bench, and present the results obtained during the on-sky tests.

\subsection{The PAPYRUS bench}

PAPYRUS is an on-sky AO platform jointly developed by the Laboratoire d'Astrophysique de Marseille (LAM) and the French aerospace lab ONERA \citep{PAPYRUS2024}. It provides a flexible AO platform designed for testing wavefront sensing concepts on sky. In this work, it served as the experimental platform for integrating and validating the IR-umPWFS.

\subsubsection{Bench description}
\label{sec:papyrus_bench_description}

\begin{figure}
\centering
\includegraphics[width=0.5\textwidth]{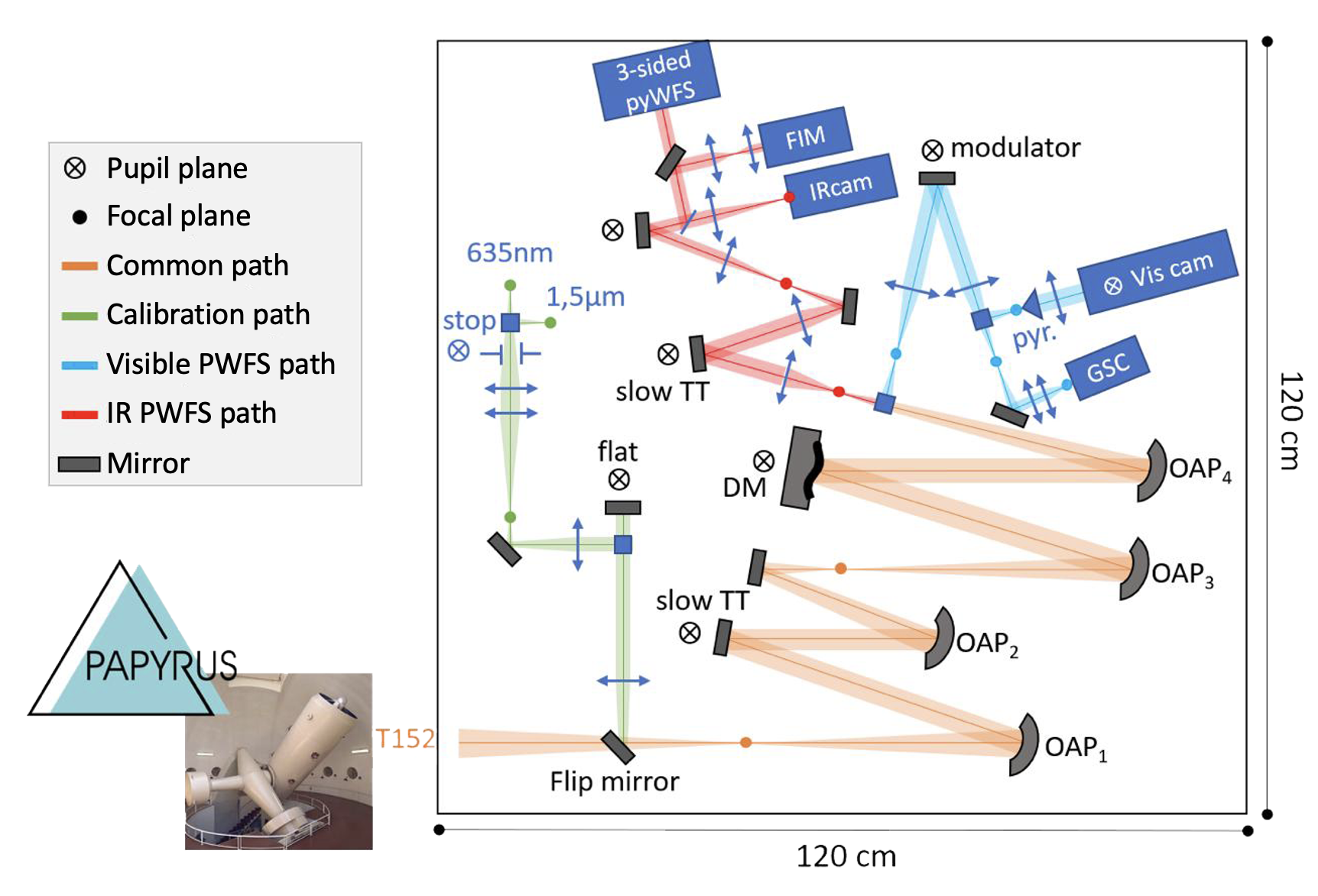}
\caption{Optical layout of the PAPYRUS adaptive optics bench, showing the common path, calibration channel, and both visible and infrared wavefront sensing arms. The system can be fed either by the telescope or by internal calibration sources via a flip mirror. The calibration channel includes visible (635\,nm) and near-infrared (1.55\,$\mu$m) light sources. The common optical path contains key elements such as a DM and slow tip-tilt (TT) mirrors for wavefront correction and offloading. A dichroic beam splitter splits the light into two channels: wavelengths shorter than 1\,$\mu$m are directed to the visible arm, while longer wavelengths are transmitted to the infrared arm, enabling simultaneous dual-band operation. The visible arm hosts a modulated 4-sided PWFS and an OCAM2K camera. The infrared arm includes the 3-sided pyramid wavefront sensor coupled to a C-RED3 detector for wavefront sensing and a C-RED2 camera for focal-plane imaging. (Credit: Angelie Alagao/LAM).}
\label{fig:papyrus_optical_layout}
\end{figure}

The PAPYRUS bench is an AO system installed on the 1.52-m telescope at the OHP. The optical layout of the bench is shown in Fig.~\ref{fig:papyrus_optical_layout}. The visible arm hosts the VIS-mPWFS operating over $\lambda_{\mathrm{WFS}} = 400$--$950$\,nm. Its main components are an FLI OCAM2K electron-multiplying charge-coupled device camera ($240 \times 240$ pixels), a 4-sided glass pyramid optic, and a tip-tilt modulator used to provide modulation for the PWFS. An ALPAO DM241 deformable mirror, comprising 241 actuators, is located in the common optical path \citep{PAPYRUS2024}. A dichroic beam splitter directs wavelengths shorter than $1000\,\mathrm{nm}$ to the visible arm, while transmitting longer wavelengths to the infrared arm, enabling simultaneous visible and infrared operation.

As part of this work, we installed an infrared arm dedicated to wavefront sensing using an IR-umPWFS operating in the near-infrared ($\lambda_{\mathrm{WFS}} =$ 1500–1700\,nm). This infrared channel uses an FLI C-RED3 camera ($640 \times 512$ pixels) for wavefront sensing in combination with the 3-sided PWFS developed in this work. During our tests, the C-RED3 was not cooled, but this was acceptable given the short exposure times used for tests. The detector had a relatively high readout noise of $30\,\mathrm{e}^-$, and the use of a 3-sided PWFS provides an additional advantage by reducing the number of pixels that must be read out, improving the effective signal-to-noise ratio of the wavefront measurements. The camera was operated in high-gain mode, and a reduced window size of $160 \times 128$ pixels was used to increase the maximum frame rate to 4\,kHz. For PSF characterization in the near-infrared (1000--1500\,nm, with an effective wavelength of $\lambda_{\mathrm{PSF}} = 1400$\,nm), an FLI C-RED2 camera ($640 \times 512$ pixels) was used. The detector sampling was 2.69 pixels per $\lambda_{\mathrm{PSF}}/D$.

The main differences between the two wavefront sensing arms are summarized in Table~\ref{tab:arm_differences}. As discussed in Sect.~\ref{sec:type_of_pyramid}, the pyramid geometry is not expected to bias the comparison, the 3-sided design having been chosen to avoid the roof defect of 4-sided pyramids. For the VIS-mPWFS, the impact of the roof defect is mitigated by the $5\,\lambda/D$ modulation, which moves the PSF over a region much larger than the defect. The infrared sensing wavelength, the absence of modulation and the resulting higher loop speed are precisely the advantages of the IR-umPWFS investigated in this work: infrared operation extends the linear range sufficiently to remove the need for modulation, and without modulation the loop speed is no longer limited by the modulation mirror speed. Finally, the two arms are affected by different non-common path aberrations (NCPA), whose contribution is removed by using the corrected Strehl ratio (Sect.~\ref{sec:strehl_ee}).

\begin{table}[t]
\centering
\caption{Main differences between the two wavefront sensing arms used for the on-sky comparison.}
\label{tab:arm_differences}
\small
\setlength{\tabcolsep}{4pt}
\resizebox{\columnwidth}{!}{%
\begin{tabular}{lcc}
\toprule
\textbf{Parameter} & \textbf{IR-umPWFS} & \textbf{VIS-mPWFS} \\
\midrule
Sensing band $\lambda_{\mathrm{WFS}}$ [nm] & 1500--1700 & 400--950 \\
Pyramid & 3-sided & 4-sided \\
Roof defect & $< 4\,\mu$m & $\sim31 \times 12\,\mu$m \\
Modulation & $0\,\lambda/D$ & $5\,\lambda/D$ \\
Loop speed [Hz] & 1000 & 500 \\
NCPA & trefoil, $\sim$75\,nm RMS & astigmatism, $\sim$100\,nm RMS \\
\bottomrule
\end{tabular}%
}
\end{table}

\subsubsection{Control architecture}

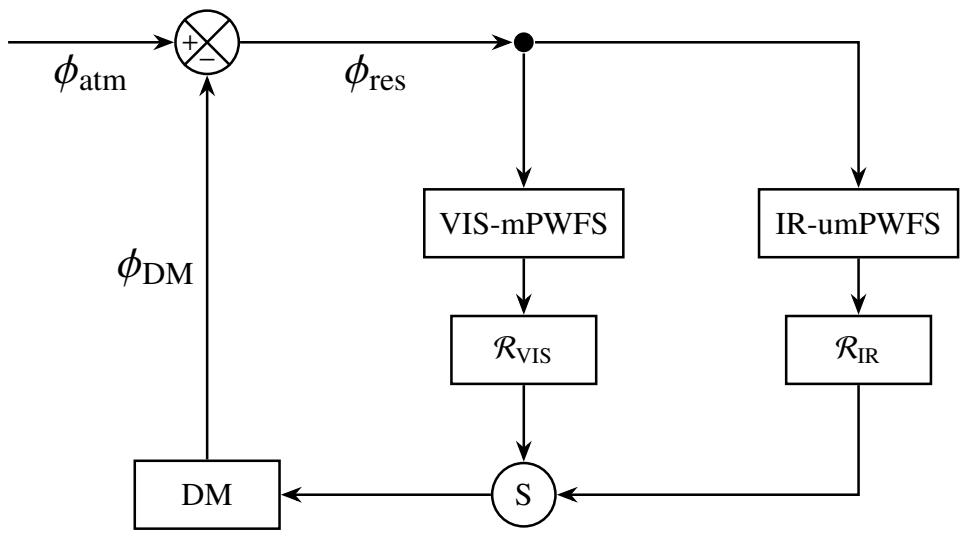
\begin{figure}[htbp]
\centering
\resizebox{0.7\columnwidth}{!}{%
\begin{tikzpicture}[
    >=Stealth,
    thick,
    block/.style={draw, minimum width=2.2cm, minimum height=0.75cm},
    smallblock/.style={draw, minimum width=1.6cm, minimum height=0.75cm},
    sum/.style={circle, draw, minimum size=0.7cm, inner sep=0pt},
    dot/.style={circle, fill, inner sep=2.2pt},
    phistyle/.style={font=\Large}
]

\node[sum] (s1) at (0,0) {};
\draw (s1.north west) -- (s1.south east);
\draw (s1.north east) -- (s1.south west);
\node at ([xshift=5pt]s1.west) {\small $+$};
\node at ([yshift=5pt]s1.south) {\small $-$};
\node[dot] (split) at (3.5,0) {};

\node[block] (vis) at (3.5,-2.0) {VIS-mPWFS};
\node[smallblock] (rmvis) at (3.5,-3.4) {$\mathcal{R}_{\mathrm{VIS}}$};

\node[block] (ir) at (7.2,-2.0) {IR-umPWFS};
\node[smallblock] (rmir) at (7.2,-3.4) {$\mathcal{R}_{\mathrm{IR}}$};

\node[sum] (s2) at (3.5,-5.0) {S};
\node[smallblock] (dm) at (0,-5.0) {DM};

\draw[->] (-2.2,0) -- node[below, phistyle] {$\phi_{\mathrm{atm}}$} (s1);
\draw[->] (s1) -- node[below, phistyle] {$\phi_{\mathrm{res}}$} (split);

\draw[->] (split) -- (vis);
\draw[->] (split) -| (ir);

\draw[->] (vis) -- (rmvis);
\draw[->] (rmvis) -- (s2);

\draw[->] (ir) -- (rmir);
\draw[->] (rmir) |- (s2);

\draw[->] (s2) -- (dm);

\draw[->] (dm.north) -- node[left, phistyle] {$\phi_{\mathrm{DM}}$} (s1.south);

\end{tikzpicture}%
}

\caption{Schematic control architecture of the system. The VIS-mPWFS and IR-umPWFS provide independent measurement paths, each with its corresponding reconstruction matrix. The node labeled S represents a switch that determines which control path is active and drives the DM.}

\label{fig:control_scheme}
\end{figure}

Real-time control was provided by the DAO RTC \citep{barr_2025_17264152}. As part of this work, we developed the necessary DAO software interface to integrate the infrared arm into the PAPYRUS control architecture. Because the AO lab bench used the same DAO framework, much of the software developed and validated during the bench experiments could be directly reused on the PAPYRUS system, allowing integration within a few hours. The VIS-mPWFS was operated using the PAPYRUS RTC implementation, while the IR-umPWFS was operated through our control pipeline validated on the bench in Geneva (Sect.~\ref{sec:closed_loop_tests}). The control architecture was designed such that a simple software switch allowed either the VIS-mPWFS or the IR-umPWFS to control the DM. Both systems functioned as independent single-stage control loops, with only one wavefront sensor actively driving the DM at any given time. The control scheme is shown in Fig.~\ref{fig:control_scheme}. Data from both wavefront sensor streams were acquired simultaneously, irrespective of which wavefront sensor was driving the DM.

\subsubsection{On-sky commissioning}
We had two observing runs of three nights each allocated on the PAPYRUS bench in July and September 2025 for integration and testing of the IR-umPWFS.

During the first observing run, we installed the optical components of the infrared arm and tested the control software developed for the system. The software integration with the PAPYRUS control system was successful and operated reliably from the first day. The remaining nights of the run were devoted to on-sky testing. The loop could be closed during this first run, but alignment issues in the infrared arm prevented us from recording useful data. During the second run, the system was successfully re-aligned. The first day was then dedicated to calibrating and testing the system, as well as integrating software updates. The following three nights were then used for on-sky observations and performance testing of the IR-umPWFS. The details of these observations are provided in Appendix~\ref{sec:appendix_on_sky_observations}.
\subsection{Calibrations}

The calibrations were performed using an infrared light source at 1550\,nm. We used the KL modes of PAPYRUS, including tip and tilt, both for calibration and for controlling the AO system. The maximum number of KL modes used at PAPYRUS is 195. 

At PAPYRUS, daytime calibrations are performed using an unobstructed pupil mask, i.e., without central obscuration or spiders. Therefore, diffraction effects caused by telescope obstructions are not included in the calibration measurements, which may slightly modify the response of the PWFS compared to nighttime operations.

\subsubsection{Reference image \& NCPA}
\label{sec:reference_image_ncpa}

\begin{figure}
\centering
\includegraphics[width=0.8\columnwidth]{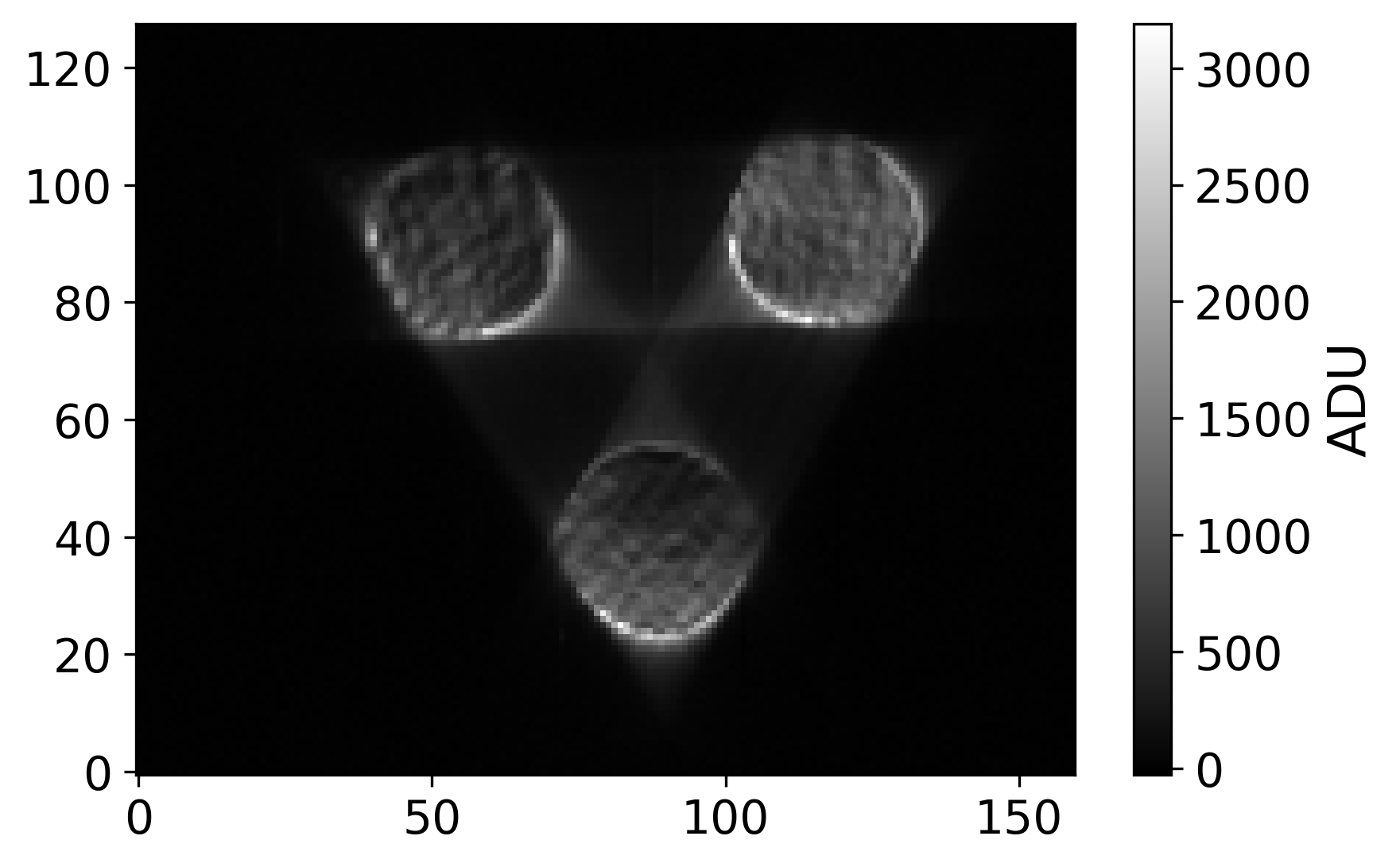}
\caption{Reference image for the IR-umPWFS obtained on the PAPYRUS bench.}
\label{fig:reference_image}
\end{figure}

The reference image corresponds to the wavefront sensor response to a flat wavefront in an ideal case like in simulations. In practice, however, the reference image defines the wavefront toward which the control loop converges, and it may include NCPA. The reference image is shown in Fig.~\ref{fig:reference_image}. 

In our case, we used the valid pixel mask as a synthetic reference (the reason for this choice is described in Sect.~\ref{sec:interaction_matrix_and_dm_flat}), which however does not correspond to a flat wavefront. This approximation generates a trefoil NCPA of $75\,\mathrm{nm}$ RMS, which is observed on the PSF camera. We did not have time to implement a correction algorithm.

\begin{figure*}
\centering
\includegraphics[width=\textwidth]{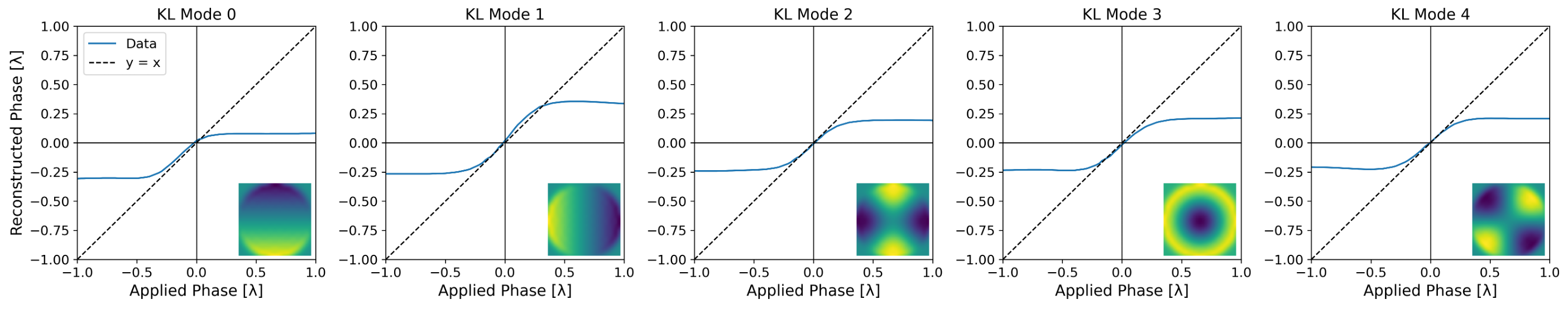}
\caption{Linearity curves of the first five KL modes at the PAPYRUS bench. The dashed line indicates the ideal linear response (y = x). The first two modes are tip-tilt. The corresponding KL mode phase is shown in the lower-right corner of each plot.}
\label{fig:linearity}
\end{figure*}

\subsubsection{Interaction matrix and DM flat}
\label{sec:interaction_matrix_and_dm_flat}

The interaction matrix was obtained through a modal calibration using KL modes. We kept the amplitude of simulations $a = 0.05$ (KL unit), and after a wait time of 50\,ms, the corresponding PWFS images were recorded. For each mode, the 100 push-pull responses were combined to average out the impact of internal turbulence.
Because the  optical quality of the infrared arm was not excellent, the calibration was obtained iteratively, in order to calibrate the system as close as possible to the reference phase. The procedure begins with an initial calibration, followed by closing the AO loop on the internal calibration source and recording the DM commands that minimize the wavefront error measured by the PWFS. During the initial calibration, the loop is closed using only the first two KL modes (tip-tilt modes) to center the beam on the sensor. The resulting DM commands are then applied as a fixed offset, and a new calibration is performed. In subsequent iterations, the number of controlled modes is progressively increased. This process is repeated until the PWFS measurements converge to a uniform intensity distribution among the pupil images. The resulting set of DM commands defines the final DM flat. The reference image, valid-pixel mask, and interaction matrix are all measured with the DM set to this flat configuration. All subsequent correction commands are applied relative to this reference state.

\subsubsection{Linearity curves}

The linearity curves of each KL mode were evaluated by applying a range of amplitudes $a_{\mathrm{applied}}$ to a given mode $\phi_i$ and recording the corresponding PWFS slopes, $S(a_{\mathrm{applied}}\phi_i)$. The amplitude measured by the PWFS, $a_{\mathrm{measured}}$, was then obtained by extracting the $i$-th component of the reconstructed modal coefficients:

\begin{equation}
a_{\mathrm{measured}} 
= \left[ \mathcal{R} \, S(a_{\mathrm{applied}}\phi_i) \right]_i
\end{equation}

\noindent where $\mathcal{R}$ is the reconstruction matrix (Sect.~\ref{sec:simulation_description}).

Plotting the applied amplitude against the reconstructed amplitude  for a particular mode yields the linearity curve for that mode. The linearity curves for the first five KL modes are shown in Fig.~\ref{fig:linearity}. For most modes, the linearity curves are well centered around zero and follow the expected linear response. However, KL mode 0 (tip-tilt) exhibits a noticeable offset, indicating the presence of a residual tip-tilt offset in the system. A slight asymmetry between the shape of responses to negative and positive amplitudes is also observed for some modes, while small phase offsets are seen in other modes. This could be due to various reasons including the relative orientation of the 3-sided pyramid with respect to the optical axis, as discussed in Sect.~\ref{sec:linearity_curves_sim_bench}, or the presence of small modal phase offsets in the system.

\subsection{Observations}

We acquired a series of open-loop and closed-loop observations with the  IR-umPWFS over three nights of on-sky testing. We performed several consecutive observations with both the  IR-umPWFS and the VIS-mPWFS to enable a side-by-side comparison of their performance. The typical observing sequence consisted of first observing with the IR-umPWFS, followed by the VIS-mPWFS, and then taking open-loop measurements. The AO system was operated using a simple integrator controller with the leak term fixed at 0.01, and the loop gain was optimized for each of the observations. 

As part of the telemetry stream, we recorded the DM command vectors, PWFS images and slopes, residual KL modes, and the PSF cube recorded by the C-RED2 camera. Most observations were acquired with a total observation time of 5 seconds, while some 30-second and 5-minute observations were taken. Longer observations could not be obtained because the DM offloading system was not working. In this section, we describe the processing of the acquired data; the processing of the PSF images is described in Appendix~\ref{sec:appendix_processing_psf_images}.

\begin{figure}
\centering
\includegraphics[width=\columnwidth]{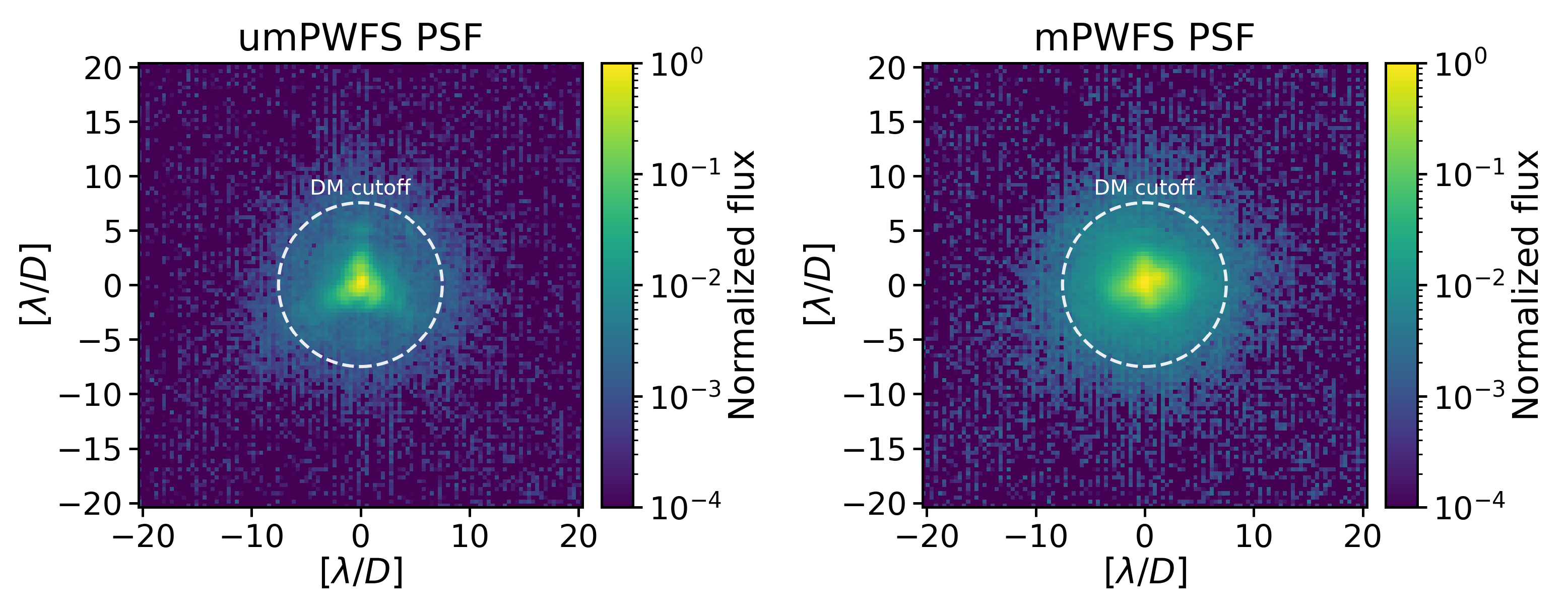}
\caption{Integrated PSFs for $5$-minute closed-loop observations with the IR-umPWFS (left) and the VIS-mPWFS (right) in log scale, at $\lambda_{\mathrm{PSF}} = 1400$\,nm. The dashed circle shows the DM cutoff or the DM correction area.}
\label{fig:PSFS_5min_obs}
\end{figure}

\subsubsection{Encircled energy}
We computed the encircled energy (EE) within a circular aperture of radius $1\,\lambda_{\mathrm{PSF}}/D$, centered on the center of mass of the PSF. This radius corresponds to the diffraction-limited core. For a diffraction-limited PSF with the 25\% central obscuration of the PAPYRUS telescope, approximately $72\%$ of the total energy is expected to be contained within a radius of $1\,\lambda_{\mathrm{PSF}}/D$. Measuring the EE within this region therefore provides a direct estimate of how much light is concentrated in the PSF core. 

\subsubsection{Strehl ratio}
\label{sec:strehl_ee}
We computed the Strehl ratio using the optical transfer function (OTF) method \citep{Lewis2004}. The observed PSF was compared to a diffraction-limited reference PSF generated with MAOPPY \citep{Fetick2019}:
\begin{equation}
S = \frac{\sum\limits_{f} \mathrm{OTF}_{\mathrm{obs}}(f)}{\sum\limits_{f} \mathrm{OTF}_{\mathrm{diff}}(f)}.
\end{equation}
\noindent where the sum is carried out over spatial frequencies f up to the telescope cutoff frequency.

The corrected Strehl ratio was estimated after removing the contribution of known static aberrations (described in Sect.~\ref{sec:reference_image_ncpa}). The Strehl ratio measured from the PSF was first converted into an equivalent WFE using the Mar\'echal approximation,
\begin{equation}
S \approx \exp\left[-\left(\frac{2\pi \sigma_{\mathrm{WFE}}}{\lambda}\right)^2\right],
\end{equation}
where $\sigma_{\mathrm{WFE}}$ is the RMS WFE and $\lambda$ is the observing wavelength. The known static aberration contribution was then subtracted from the total WFE: 
\begin{equation}
\sigma_{\mathrm{corr}} = \sqrt{\sigma_{\mathrm{total}}^2 - \sigma_{\mathrm{static}}^2},
\end{equation}
\noindent where $\sigma_{\mathrm{total}}$ is the total measured WFE  and $\sigma_{\mathrm{static}}$ is the known static aberration contribution. Finally, the corrected WFE was converted back into a Strehl ratio using the same approximation.

\subsection{On-sky performance}

\subsubsection{IR-umPWFS 1000\,Hz vs.\ VIS-mPWFS 500\,Hz}
\label{sec:ir_umpwfs_1000hz_vs_vis_mpwfs_500hz}

We present the observed performance of the two wavefront sensors in their default configurations: the IR-umPWFS operating at $1000\,\mathrm{Hz}$ and the VIS-mPWFS at $500\,\mathrm{Hz}$. Due to the absence of atmospheric dispersion correction, a tip-tilt offset of $\sim3\,\lambda_{\mathrm{PSF}}/D$ at $\lambda_{\mathrm{PSF}} = 1400\,\mathrm{nm}$ was present between the WFSs. Although beyond the linear regime, both sensors compensated it effectively, allowing stable operation and switching, demonstrating robustness to large tip-tilt offsets and potential vibrations.

\begin{figure}
\centering
\includegraphics[width=0.8\columnwidth]{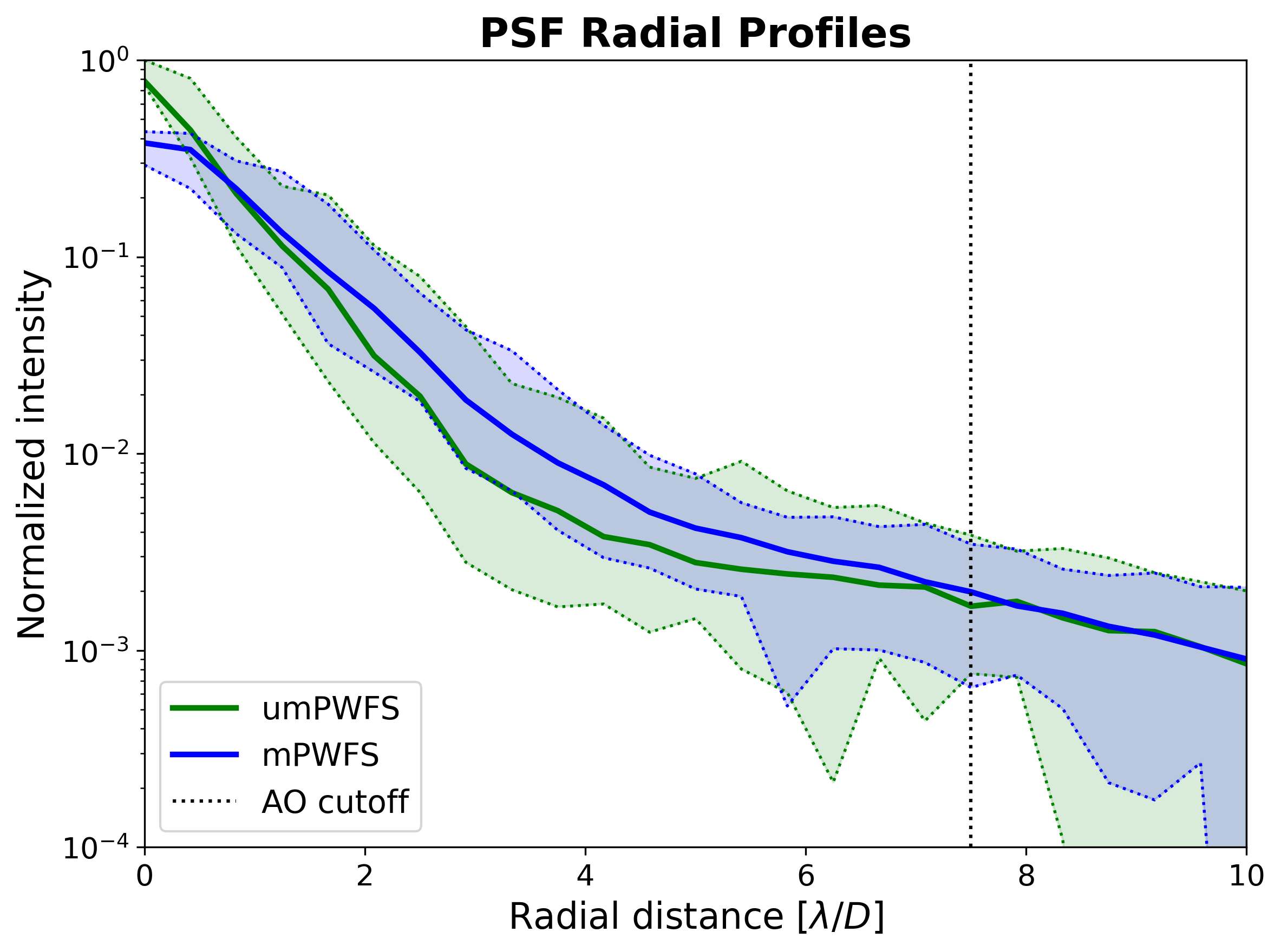}
\caption{Median radial profiles of the normalized PSF for the IR-umPWFS (green) and VIS-mPWFS (blue), expressed in units of $\lambda_{\mathrm{PSF}}/D$ at $\lambda_{\mathrm{PSF}} = 1400\,\mathrm{nm}$. Solid lines show the median radial profile, while shaded regions and dotted lines indicate the minimum-maximum envelope at each radius. The vertical dashed line marks the DM cutoff frequency.}
\label{fig:PSFS_radial_profile}
\end{figure}

Figure~\ref{fig:PSFS_5min_obs} presents typical PSFs for each observation. Figure~\ref{fig:PSFS_radial_profile} shows the corresponding median radial profiles. The DM correction region is visible in the IR-umPWFS PSF, confirming correction up to $f = 7.5\,\lambda_{\mathrm{PSF}}/D$. A lower halo level is observed for the IR-umPWFS, indicating improved residual WFE correction. NCPA are clearly present, with a trefoil aberration in the IR-umPWFS PSF (Sect.~\ref{sec:reference_image_ncpa}), while a $\sim100\,\mathrm{nm}$ RMS astigmatism is present in the VIS-mPWFS PSF, which could not be corrected during this run. As seen in Fig.~\ref{fig:PSFS_radial_profile}, at $5\,\lambda_{\mathrm{PSF}}/D$, the contrast reaches $\sim3\times10^{-3}$ (IR-umPWFS) versus $\sim7\times10^{-3}$ (VIS-mPWFS). Beyond the DM cutoff ($7.5\,\lambda_{\mathrm{PSF}}/D$), both converge to $\sim2\times10^{-3}$. Profiles beyond $10\,\lambda_{\mathrm{PSF}}/D$ are noise-dominated. Shaded regions indicate the min–max envelope, tracing optimal contrast and residual aberration contributions.

Table~\ref{tab:perf} summarizes the performance of the two arms, whose main differences are listed in Table~\ref{tab:arm_differences}. The IR-umPWFS achieves smaller FWHM, higher Strehl ratio, and EE under the same seeing conditions ($2.4''$, computed from open-loop PSF).

\begin{table}[t]
\caption{Performance comparison between the IR-umPWFS and VIS-mPWFS for a 5-minute observation.}
\label{tab:perf}
\centering
\begin{tabular}{lcc}
\hline\hline
 Parameter & IR-umPWFS & VIS-mPWFS \\
\hline
Loop speed [Hz]          & 1000   & 500    \\
Peak flux [ADU]          & 450.36 & 230.59 \\
EE [\%]                  & 34.03  & 24.14  \\
FWHM [$\lambda_{\mathrm{PSF}}/D$]       & 1.47   & 2.38   \\
Strehl [\%]              & 28.06  & 12.68  \\
Corrected Strehl [\%]    & 31.39  & 15.48  \\
Seeing [arcsec]          & $2.4''$ & $2.4''$ \\
\hline
\end{tabular}
\end{table}

As discussed in Sect.~\ref{sec:papyrus_bench_description} (Table~\ref{tab:arm_differences}), this comparison is a system-level one. Simulations of the IR-umPWFS alone (Sect.~\ref{sec:strehl_comparison_sims}) show that increasing the loop speed from $500\,\mathrm{Hz}$ to $1\,\mathrm{kHz}$ substantially improves performance, suggesting that the higher loop speed is a major contributor to the observed gain, although the on-sky data do not allow its contribution to be separated from that of infrared, unmodulated sensing. The higher loop speed enables improved temporal correction, resulting in sharper PSFs and higher Strehl.
We conducted multiple observations comparing both systems (see Table~\ref{tab:3pwfs_1000_4pwfs_500}). Gains and controlled modes were optimized for each case.

\begin{figure}
   \centering
   \includegraphics[width=0.8\columnwidth]{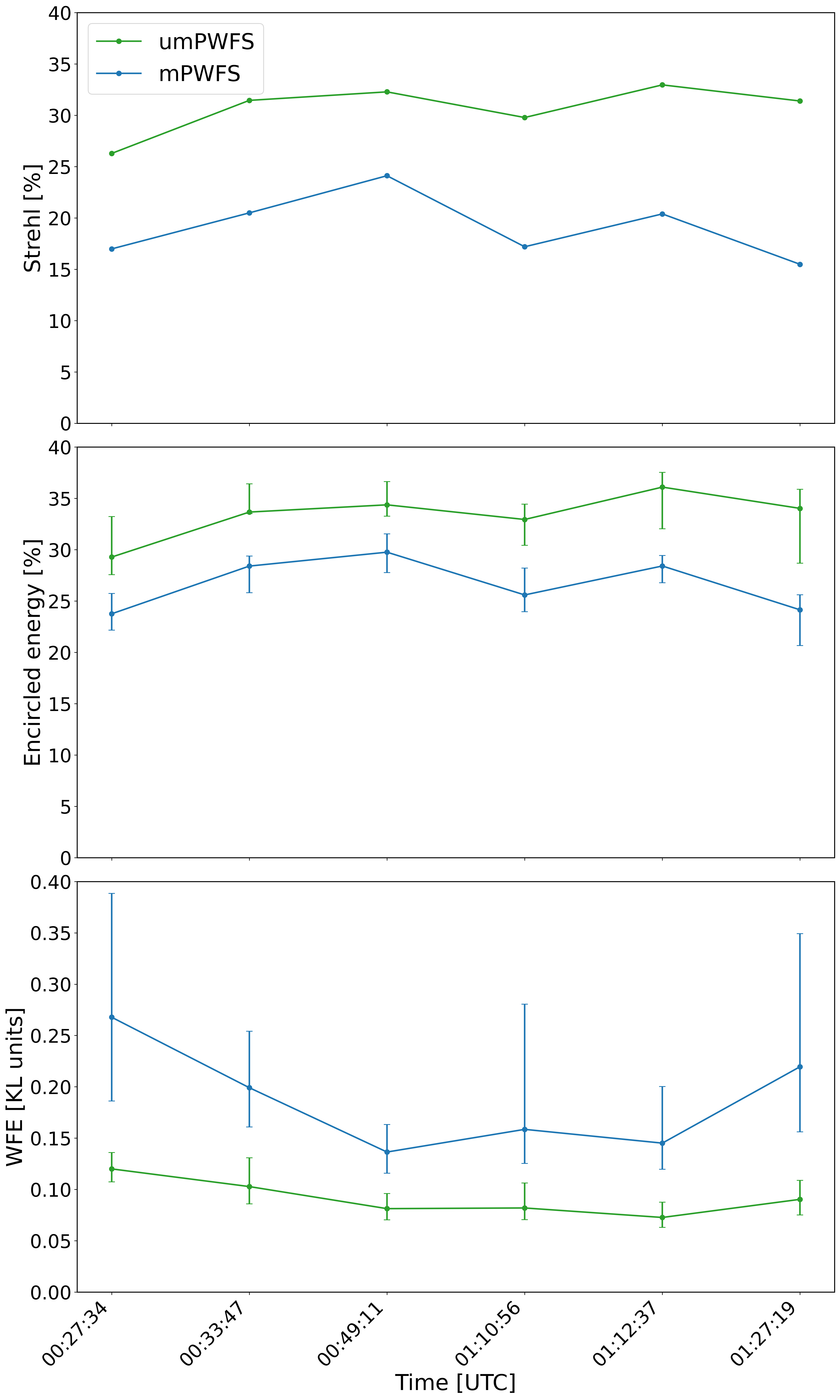}
   \caption{Comparison of the Strehl (top), EE (middle), and RMS WFE (bottom) for several observations with the IR-umPWFS (green; $1000\,\mathrm{Hz}$) and VIS-mPWFS (blue; $500\,\mathrm{Hz}$). Error bars represent the $16^\mathrm{th}$--$84^\mathrm{th}$ percentile range (equivalent to a $1\sigma$ range for a Gaussian distribution). Time is given in UTC, and each point corresponds to a closed-loop observation with durations ranging from $5\,\mathrm{s}$ to $5\,\mathrm{min}$. Each time stamp corresponds to the start of the IR-umPWFS observation of a pair; the paired VIS-mPWFS observation was acquired within a few minutes (see Table~\ref{tab:3pwfs_1000_4pwfs_500} for the exact times).}
   \label{fig:EE_VS_TIME}
\end{figure}

Figure~\ref{fig:EE_VS_TIME} compares Strehl, EE, and RMS WFE across datasets. The IR-umPWFS consistently outperforms the VIS-mPWFS, with average gains of $\sim1.26$ in EE (range:
1.15–1.41) and $\sim1.63$ in Strehl (range: 1.34–2.03). This reflects both improved light concentration and reduced residual WFE, benefiting applications such as fiber injection. The gain is robust across all observations.

A similar trend is observed in WFE: reduced from $\sim0.19$ to $\sim0.09$ KL units (factor $\sim2.03$; range: 1.68–2.43), with smaller dispersion, indicating more stable correction. This confirms that performance gains originate from reduced residual aberrations.

\begin{figure}
   \centering
   \includegraphics[width=\columnwidth]{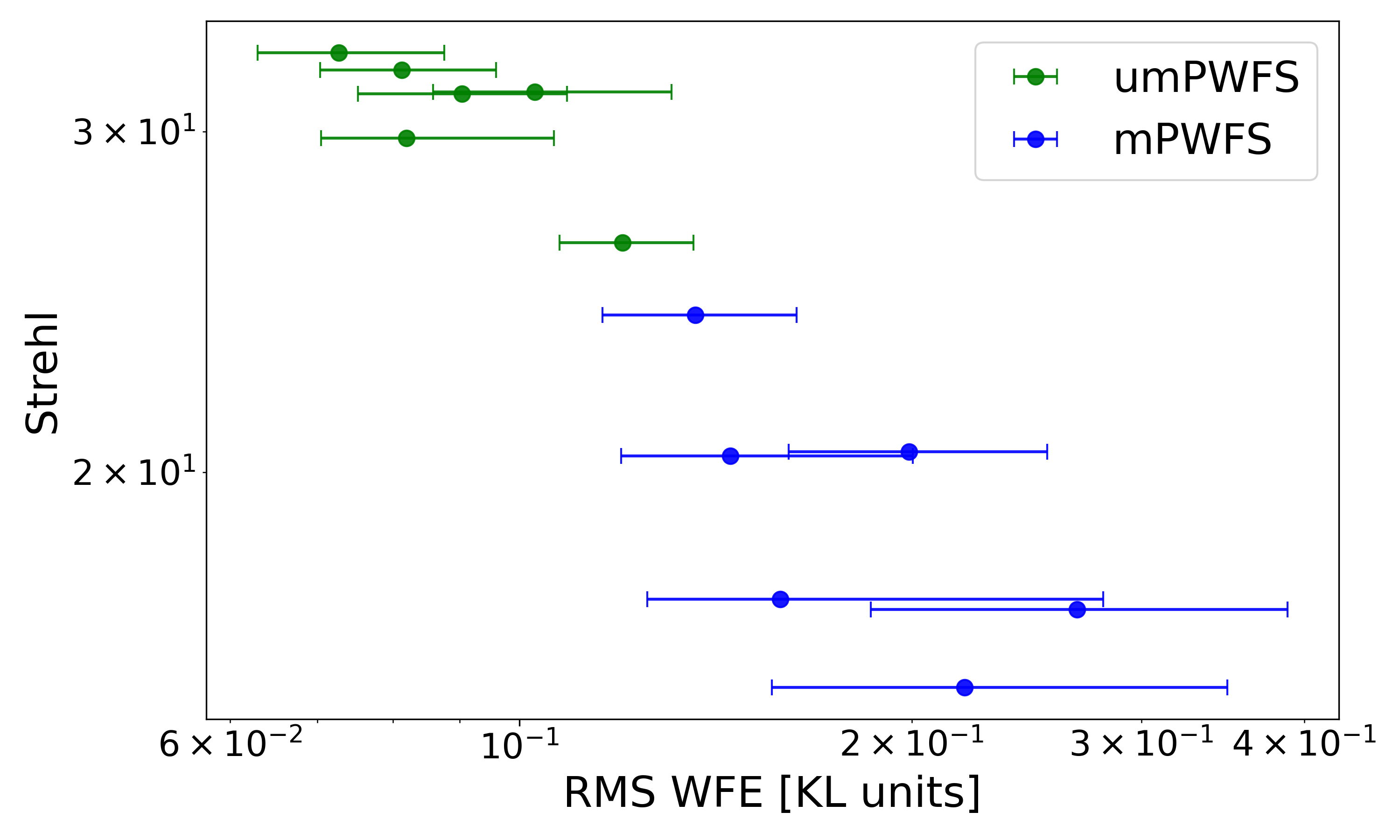}
   \caption{Strehl vs RMS WFE comparison for several observations with IR-umPWFS (green; $1000\,\mathrm{Hz}$) and VIS-mPWFS (blue; $500\,\mathrm{Hz}$). Error bars  correspond to the $16^\mathrm{th}$--$84^\mathrm{th}$ percentile range (this interval is equivalent to a $1\sigma$ range for a Gaussian distribution, while remaining more robust for non-Gaussian data) of the RMS WFE.}
   \label{fig:Strehl_vs_RMS}
\end{figure}

In Fig.~\ref{fig:Strehl_vs_RMS}, the Strehl–RMS relation appears linear in log–log space, consistent with the Maréchal approximation. The good agreement between the Strehl values derived from the PSF and the RMS WFE estimated from PWFS measurements demonstrates the consistency of the two metrics.

We also computed the tip-tilt jitter as a measure of image stability. As shown in Fig.~\ref{fig:tip-tilt_jitter_vs_time} in Appendix~\ref{sec:tip_tilt_jitter}, the jitter levels are comparable between the IR-umPWFS and VIS-mPWFS. Overall, the tip-tilt jitter remains below one pixel for all observations, despite observation sequences ranging from a few seconds to several minutes, indicating stable closed-loop performance in both configurations.

\subsubsection{IR-umPWFS 500\,Hz vs.\ VIS-mPWFS 500\,Hz}
\label{sec:ir500_vs_vis500}

We performed observations to compare the closed-loop performance of the IR-umPWFS and VIS-mPWFS at the same loop speed of 500\,Hz. The corresponding observation parameters and performance metrics are summarized in Table~\ref{tab:3pwfs_4pwfs_500hz}. For each dataset, the loop gain and the number of controlled modes were optimized to maximize performance. At equal loop speed, the comparison between the two wavefront sensing schemes is inconclusive. While the IR-umPWFS achieves higher Strehl ratio and EE in one case, in the other two observations lower Strehl ratios are obtained with the IR-umPWFS, which can be attributed to instabilities caused by brief openings of the control loop during the acquisition, leading to a degradation of the average PSF. This complicates a direct comparison between datasets.

Here we did not observe the performance gain predicted by simulations for infrared operation \citep{Bond2018} for the IR-umPWFS. This discrepancy is likely due to the dominance of temporal lag errors in the error budget at 500\,Hz, which limits the achievable correction and may mask the potential advantage of the IR-umPWFS operated at longer wavelengths.

\subsubsection{IR-umPWFS 1000\,Hz vs. 500\,Hz}

We also compared the closed-loop performance of the IR-umPWFS at two different loop speeds, 500\,Hz and 1000\,Hz. The corresponding parameters and performance metrics are listed in Table~\ref{tab:3pwfs_500_1000}.

In principle, increasing the loop speed is expected to improve performance, particularly in terms of Strehl ratio, with simulations predicting gains of 10--50\% depending on the turbulence conditions as shown in Fig.~\ref{fig:Simulations_at_delay1}. However, this trend is not  observed in our on-sky dataset. The comparison is limited by the small number of observations, comprising only one dataset at 1000\,Hz and three at 500\,Hz. In addition, two of the 500\,Hz observations are affected by instabilities caused by brief openings of the control loop during acquisition. These transient events degraded the average PSF, increased the RMS wavefront error, and complicated a direct comparison between datasets. Consequently, no firm conclusion can be drawn regarding the impact of loop speed based on this limited dataset.

\subsection{Comparison with simulations}

\subsubsection{Rejection transfer function comparison}

We analyze the rejection transfer function (RTF), which characterizes the frequency-dependent attenuation of disturbances by the closed-loop system.

The experimental RTF is estimated from the telemetry data. For each mode, the temporal power spectral densities (PSDs) of the pseudo-open-loop (POL) signal, $\Phi_{\mathrm{POL}}(f)$, and the closed-loop residual signal, $\Phi_{\mathrm{res}}(f)$, are computed. The empirical rejection transfer function is then obtained as

\begin{equation}
\mathrm{RTF}_{\mathrm{emp}}(f) = \frac{\Phi^{\mathrm{res}}(f)}{\Phi^{\mathrm{POL}}(f)}
\end{equation}

The theoretical rejection transfer function is computed from a model of the AO loop as the sensitivity function
\begin{equation}
\mathrm{RTF}_{\mathrm{th}}(f) = \frac{1}{1 + G(f)K(f)},
\end{equation}
where $K(f)$ represents the controller, modeled here as an integrator with fixed gain, and $G(f)$ describes the system response, including the loop delay \citep{GendronLena1994, Ogata2010}.

\begin{figure}

\centering
\includegraphics[width=\columnwidth]{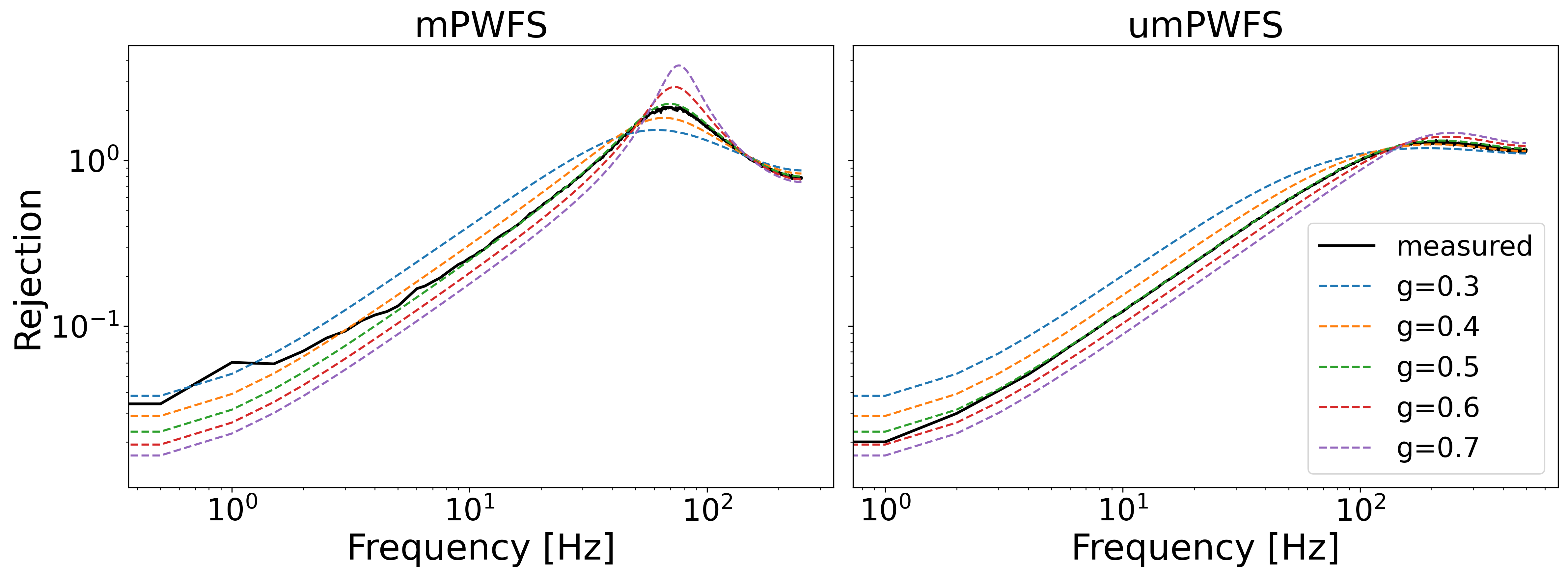}
\caption{Comparison of measured and theoretical rejection transfer functions (RTFs) for the VIS-mPWFS (left) and IR-umPWFS (right) for KL mode 5. The measured RTF (black) is shown alongside theoretical models for different loop gains (dashed curves).}
\label{fig:RTF_vs_frequency}
\end{figure}

For each observation, an RTF is computed for every KL mode, and compared to the model (Fig.~\ref{fig:RTF_vs_frequency}). We used delays of 1.2 frames for the IR-umPWFS operating at 1000\,Hz and 2 frames for the VIS-mPWFS at 500\,Hz, as estimated during operations on sky. This leads to a uniform gain of $g \simeq 0.5$ in both cases consistent with the gain applied during closed-loop operation. The RTFs further show strong rejection of low-frequency turbulence and indicate loop bandwidths of approximately 40\,Hz for the VIS-mPWFS and 100\,Hz for the IR-umPWFS.

\subsubsection{Strehl comparison}
\label{sec:strehl_comparison_sims}

We performed end-to-end simulations to investigate which atmospheric conditions could reproduce the on-sky closed-loop performance. The simulations were configured to match the flux and noise levels measured in the PWFS images during the observations. In particular, we adopted a detector dark level of $\sim 1500$ analog-to-digital units (ADU)~pixel$^{-1}$~frame$^{-1}$ and a read noise of $10$\,ADU. The total flux was set to $5$–$6 \times 10^{6}$ photons per WFS frame for an integration time of $1$\,ms to get pupil illumination levels of approximately $2000$–$3000$\,ADU~pixel$^{-1}$ that matched the observed values. Photon noise was modeled assuming Poisson statistics applied to the detected signal. The simulation architecture and parameters are the same as described in Table~\ref{tab:simulation_parameters} and Sect.~\ref{sec:simulation_description}, with the addition of realistic detector and photon-noise terms.

\begin{figure}

\centering
\includegraphics[width=\columnwidth]{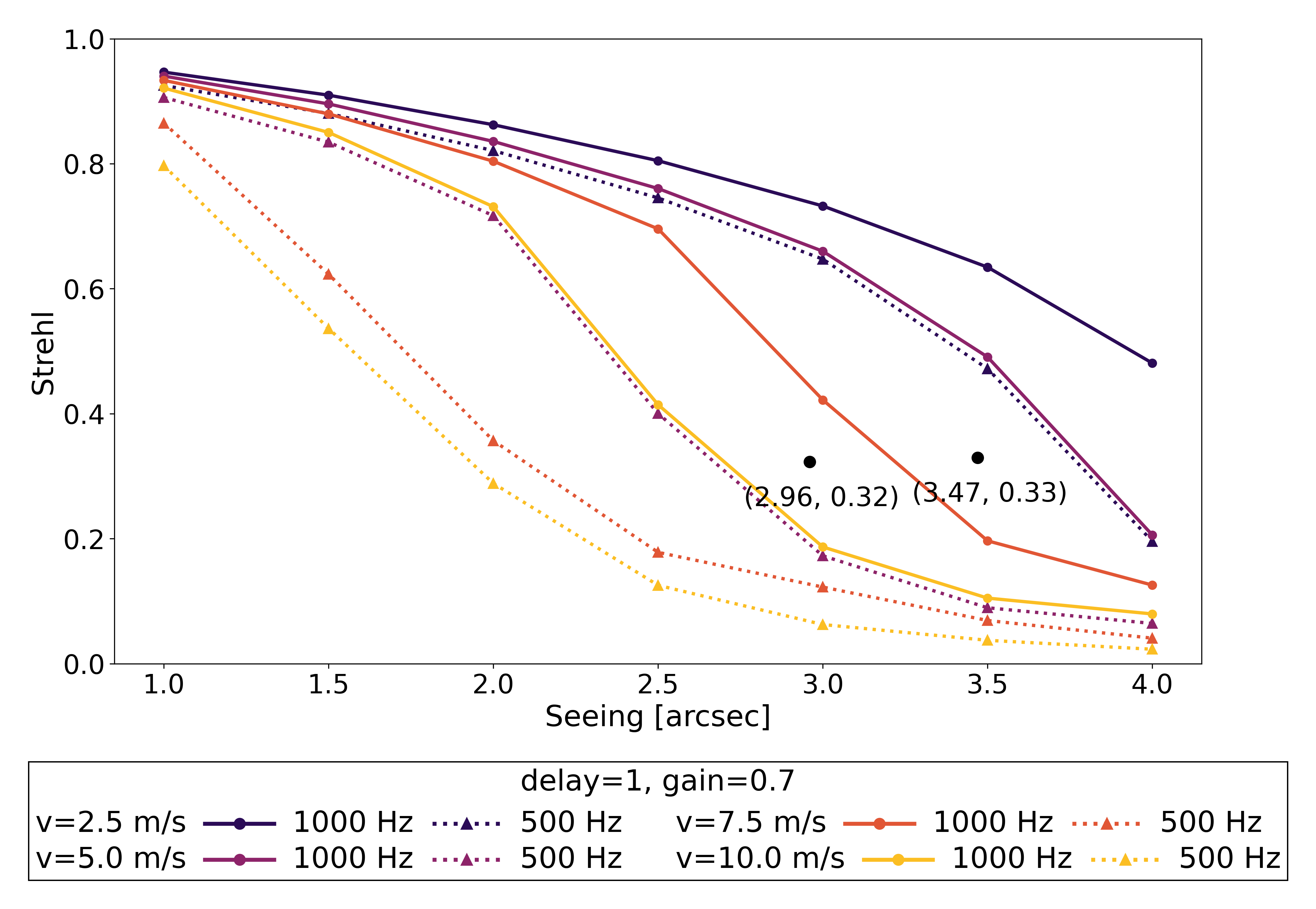}
\caption{Mean Strehl ratios from end-to-end simulations as a function of atmospheric seeing ($1''$–$4''$) and wind speed ($2.5$–$10\,\mathrm{m\,s^{-1}}$). A loop delay of 1 frame and a gain of 0.7 were adopted. The two black points indicate the on-sky measurements, with Strehl plotted at the corresponding estimated seeing conditions.}
\label{fig:Simulations_at_delay1}
\end{figure}

The on-site conditions observed during our observations were a seeing of $2$–$4''$ and wind speeds of $5$–$20\,\mathrm{m\,s^{-1}}$. To investigate their impact, we carried out a grid of simulations covering a representative range of atmospheric parameters (the seeing was varied between $1''$ and $4''$, and wind speeds between $2.5$ and $10\,\mathrm{m\,s^{-1}}$). The results are presented in terms of mean Strehl ratio in Fig.~\ref{fig:Simulations_at_delay1}. A delay of 1 frame was taken which is consistent with the latency observed during real-time operation, and a loop gain of 0.7, which yielded the best performance in the simulations. In contrast, on-sky data showed optimal gains in the range 0.3–0.5, which provided the best performance. The two black points in Fig.~\ref{fig:Simulations_at_delay1} represent on-sky measurements, with Strehl ratios plotted against the seeing estimated from open-loop PSFs acquired immediately prior to the corresponding closed-loop observations. Both measurements lie in the poor-seeing regime ($\sim3''$), with Strehl ratios of $\sim32$–$33\%$ at $\lambda_{\mathrm{PSF}} = 1400\,\mathrm{nm}$. When comparing with the simulations, these values are most consistent with wind speeds in the range $5$–$10\,\mathrm{m\,s^{-1}}$.

A clear trend is observed whereby the Strehl ratio decreases with increasing wind speed. For seeing values of $3''$–$3.5''$, as observed on sky, increasing the wind speed by a factor of two can result in a Strehl degradation of $\sim10$–$30\%$.

Increasing the loop speed from $500\,\mathrm{Hz}$ to $1000\,\mathrm{Hz}$ significantly improves performance, particularly under poor seeing and high wind speeds where temporal lag errors dominate. Similar performance is obtained for cases with the same ratio $v/f_{\mathrm{loop}}$ (e.g., $v=2.5\,\mathrm{m\,s^{-1}}$ at $500\,\mathrm{Hz}$ and $v=5.0\,\mathrm{m\,s^{-1}}$ at $1000\,\mathrm{Hz}$), highlighting the dominant role of temporal lag effects in determining AO performance. These results demonstrate that increasing the loop speed is an effective way to mitigate temporal errors and maintain high Strehl ratios under realistic observing conditions.

\section{Conclusions}
\label{sec:conclusions}

The goal of this work was to assess the feasibility and performance of the IR-umPWFS as a wavefront-sensing solution for future high-contrast, high-dispersion instruments, in particular VLT/RISTRETTO.  In this context, the main objective was to evaluate whether an umPWFS can be operated in the infrared. PAPYRUS provided the opportunity to evaluate its performance against a benchmark system routinely operated with a VIS-mPWFS.

We demonstrated through simulations that the IR-umPWFS can operate reliably under a wide range of atmospheric conditions, achieving Strehl ratios of up to $\sim60\%$ at $\lambda_{\mathrm{PSF}} = 1400\,\mathrm{nm}$ for seeing conditions up to $4''$ under PAPYRUS conditions. Working in the infrared provides several advantages, including improved linearity, a reduced optical gain effect (i.e., optical gains closer to unity and preserved sensitivity in closed loop), and lower cross-talk. A prototype of the pyramid was developed and tested in the laboratory, where the results showed excellent agreement with the simulations. This consistency validates both the high optical quality of the pyramid and our theoretical understanding and implementation of the system. 

On sky, we successfully closed the adaptive optics loop with the IR-umPWFS on the PAPYRUS bench and achieved robust operation. The on-sky commissioning highlighted several practical challenges, in particular the definition of an appropriate reference image and the determination of the DM flat. In our case, the initial use of a synthetic flat reference based on the valid-pixel mask enabled the loop to converge and allowed us to determine a suitable DM flat iteratively. This procedure was essential for reaching a stable operating point on sky. However, it also introduced a residual trefoil aberration of approximately $75\,\mathrm{nm}$ RMS, which was visible in the closed-loop PSFs. Future work will focus on improving the calibration strategy, tests with an appropriate reference image, as well as the implementation of strategies to mitigate trefoil aberrations. Linearity curves provide a powerful diagnostic tool to assess the quality of the calibration, identify phase offsets in individual modes, and determine the true zero-phase reference of the system. These approaches will be key to mitigating residual static aberrations.

An offset of $3\,\lambda_{\mathrm{PSF}}/D$ between the two pyramids was present and had to be corrected each time the system switched between them; the IR-umPWFS was able to compensate for this offset reliably, demonstrating its capability to correct large tip-tilt errors.

The IR-umPWFS operating at 1000~Hz consistently demonstrated better
performance than the VIS-mPWFS operating at 500~Hz, with improvements
in both Strehl ratio and EE. On average, the Strehl ratio increased by a factor of $\sim1.63$, while the EE improved by a factor of $\sim1.26$, and RMS WFE reduced by a factor of $\sim2.03$. This is a system-level comparison, as the two arms also differ in sensing wavelength, pyramid geometry and roof defect, modulation, loop speed and NCPA (Table~\ref{tab:arm_differences}). However, the 3-sided and 4-sided pyramids show very similar performance in simulations \citep{Shinde2024}, infrared operation is what allows modulation to be removed, the absence of modulation is precisely what allows the higher loop speed, and the NCPA contribution is removed from the corrected Strehl ratios. Simulations of the IR-umPWFS indicate that the higher loop speed alone can account for a large part of this gain (Sect.~\ref{sec:strehl_comparison_sims}), with infrared operation potentially providing an additional contribution; the on-sky data, however, do not allow these contributions to be separated. Further tests with both sensors operating at the same speed are needed to confirm the gain of operation at infrared wavelengths.

Since the seeing conditions at OHP correspond to $D/r_0 \sim 17$ at $\lambda_{\mathrm{PSF}} = 1400\,\mathrm{nm}$, which is comparable to typical VLT conditions of $D/r_0 \sim 20$ at the same wavelength, the performance measured on the PAPYRUS bench is representative of what can be expected at the VLT. These results indicate that the IR-umPWFS is a promising candidate for the VLT/RISTRETTO instrument.

The current system performance was limited by the speed of the DM, restricting operation to approximately 1000\,Hz. Further tests with a faster DM are therefore required to fully demonstrate the performance gains achievable at higher loop speeds ($> 2$\,kHz). In addition, the integration of more advanced control algorithms, such as DD4AO \citep{Dinis2024} and PO4AO \citep{Nousiainen2022}, is expected to further enhance the overall performance and stability of the system.

\begin{acknowledgements}
This work has been carried out within the framework of the National Centre of Competence in Research PlanetS supported by the Swiss National Science Foundation under grant 51NF40\_205606. The authors acknowledge the financial support of the SNSF.  We also extend our sincere thanks to the PAPYRUS team for providing the platform used to conduct our tests. In addition, we thank the OHP for granting access to their telescope facilities for on-sky testing.
\end{acknowledgements}

\bibliographystyle{aa}
\bibliography{references}

\clearpage
\onecolumn
\begin{appendix}
\nolinenumbers
\section{Data tables}
\label{sec:appendix_on_sky_observations}

\begin{table*}[!htbp]
\centering
\caption{Observation parameters and performance metrics for closed-loop observations with the IR-umPWFS at loop speeds of 1000\,Hz and the VIS-mPWFS at loop speeds of 500\,Hz.}
\label{tab:3pwfs_1000_4pwfs_500}

\renewcommand{\arraystretch}{1.6}

\resizebox{\textwidth}{!}{%
\begin{tabular}{l l c c c c c c c c}
\hline \hline
Date & Type & N Modes & Gain & FWHM & RMS WFE & Loop Speed & EE & Corrected Strehl & Jitter \\
{[UTC]} &  &  &  & [$\lambda_{\mathrm{PSF}}/D$] & [KL units] & [Hz] & [\%] & [\%] & [$\lambda_{\mathrm{PSF}}/D$] \\
\hline 
2025-10-02\_00-27-34 & 3-sided & 100 & 0.4 & 1.51 & $0.1200^{+0.0160}_{-0.0126}$ & 1000 & $29.28^{+3.94}_{-1.70}$ & 26.28 & 0.39 \\[1pt]
2025-10-02\_00-27-46 & 4-sided & 100 & 0.4 & 2.22 & $0.2678^{+0.1205}_{-0.0817}$ & 500 & $23.75^{+1.97}_{-1.59}$ & 16.98 & 0.25 \\[1pt]
2025-10-02\_00-33-47 & 3-sided & 195 & 0.3 & 1.45 & $0.1028^{+0.0281}_{-0.0169}$ & 1000 & $33.67^{+2.75}_{-0.14}$ & 31.45 & 0.19 \\[1pt]
2025-10-02\_00-34-00 & 4-sided & 195 & 0.3 & 2.14 & $0.1991^{+0.0550}_{-0.0382}$ & 500 & $28.40^{+0.98}_{-2.59}$ & 20.49 & 0.16 \\[1pt]
2025-10-02\_00-48-31 & 4-sided & 170 & 0.4 & 1.93 & $0.1364^{+0.0267}_{-0.0206}$ & 500 & $29.76^{+1.78}_{-1.98}$ & 24.11 & 0.12 \\[1pt]
2025-10-02\_00-49-11 & 3-sided & 170 & 0.4 & 1.41 & $0.0813^{+0.0147}_{-0.0110}$ & 1000 & $34.37^{+2.26}_{-1.10}$ & 32.28 & 0.22 \\[1pt]
2025-10-02\_01-10-56 & 3-sided & 170 & 0.4 & 1.46 & $0.0819^{+0.0244}_{-0.0115}$ & 1000 & $32.94^{+1.49}_{-2.53}$ & 29.78 & 0.16 \\[1pt]
2025-10-02\_01-11-10 & 4-sided & 170 & 0.4 & 2.22 & $0.1585^{+0.1220}_{-0.0332}$ & 500 & $25.59^{+2.62}_{-1.63}$ & 17.19 & 0.36 \\[1pt]
2025-10-02\_01-12-37 & 3-sided & 170 & 0.5 & 1.40 & $0.0727^{+0.0149}_{-0.0097}$ & 1000 & $36.10^{+1.43}_{-4.05}$ & 32.96 & 0.14 \\[1pt]
2025-10-02\_01-13-19 & 4-sided & 170 & 0.5 & 2.08 & $0.1451^{+0.0552}_{-0.0254}$ & 500 & $28.41^{+1.02}_{-1.64}$ & 20.39 & 0.27 \\[1pt]
2025-10-02\_01-27-19 & 3-sided & 170 & 0.5 & 1.47 & $0.0904^{+0.0185}_{-0.0152}$ & 1000 & $34.03^{+1.85}_{-5.34}$ & 31.39 & 0.23 \\[1pt]
2025-10-02\_01-30-43 & 4-sided & 170 & 0.5 & 2.38 & $0.2196^{+0.1297}_{-0.0634}$ & 500 & $24.14^{+1.47}_{-3.48}$ & 15.48 & 0.20 \\
\hline
\end{tabular}%
}

\renewcommand{\arraystretch}{1}

\end{table*}

\begin{table*}[h]
\centering
\caption{Observation parameters and performance metrics for closed-loop observations with the IR-umPWFS and VIS-mPWFS at a loop speed of 500\,Hz.}
\label{tab:3pwfs_4pwfs_500hz}

\renewcommand{\arraystretch}{1.6}

\resizebox{\textwidth}{!}{%
\begin{tabular}{l l c c c c c c c c}
\hline \hline
Date & Type & N Modes & Gain & FWHM & RMS WFE & Loop Speed & EE & Corrected Strehl & Jitter \\
{[UTC]} &  &  &  & [$\lambda_{\mathrm{PSF}}/D$] & [KL units] & [Hz] & [\%] & [\%] & [$\lambda_{\mathrm{PSF}}/D$] \\
\hline

2025-10-02\_00-34-00 & 4-sided & 195 & 0.3 & 2.14 &
$0.1991^{+0.0550}_{-0.0382}$ & 500 &
$28.40^{+0.98}_{-2.59}$ & 20.49 & 0.16 \\[1pt]

2025-10-02\_00-36-37 & 3-sided & 170 & 0.7 & 1.47 &
$0.0770^{+0.0169}_{-0.0106}$ & 500 &
$34.67^{+1.01}_{-1.73}$ & 31.15 & 0.09 \\[1pt]

2025-10-02\_00-36-51 & 4-sided & 170 & 0.7 & 1.87 &
$0.1379^{+0.0278}_{-0.0171}$ & 500 &
$29.80^{+0.48}_{-1.31}$ & 25.83 & 0.07 \\[1pt]

2025-10-02\_00-37-57 & 3-sided & 170 & 0.7 & 1.81 &
$0.0917^{+0.0199}_{-0.0137}$ & 500 &
$27.61^{+6.47}_{-5.83}$ & 21.33 & 0.31 \\[1pt]

2025-10-02\_00-38-21 & 4-sided & 170 & 0.7 & 1.96 &
$0.1396^{+0.0357}_{-0.0229}$ & 500 &
$30.12^{+0.37}_{-1.93}$ & 24.85 & 0.09 \\[1pt]

2025-10-02\_00-40-53 & 3-sided & 170 & 0.4 & 1.82 &
$0.1078^{+0.0196}_{-0.0151}$ & 500 &
$27.01^{+5.40}_{-7.41}$ & 20.30 & 0.37 \\

\hline
\end{tabular}
}

\renewcommand{\arraystretch}{1}

\end{table*}

\begin{table*}[h]
\centering
\caption{Observation parameters and performance metrics for closed-loop observations with the IR-umPWFS at loop speeds of 500\,Hz and 1000\,Hz.}
\label{tab:3pwfs_500_1000}

\renewcommand{\arraystretch}{1.6}

\resizebox{\textwidth}{!}{%
\begin{tabular}{l l c c c c c c c c}
\hline \hline
Date & Type & N Modes & Gain & FWHM & RMS WFE & Loop Speed & EE & Corrected Strehl & Jitter \\
{[UTC]} &  &  &  & [$\lambda_{\mathrm{PSF}}/D$] & [KL units] & [Hz] & [\%] & [\%] & [$\lambda_{\mathrm{PSF}}/D$] \\
\hline

2025-10-02\_00-33-47 & 3-sided & 195 & 0.3 & 1.45 &
$0.1028^{+0.0281}_{-0.0169}$ & 1000 &
$33.67^{+2.75}_{-0.14}$ & 31.45 & 0.19 \\[1pt]

2025-10-02\_00-36-37 & 3-sided & 170 & 0.7 & 1.47 &
$0.0770^{+0.0169}_{-0.0106}$ & 500 &
$34.67^{+1.01}_{-1.73}$ & 31.15 & 0.09 \\[1pt]

2025-10-02\_00-37-57 & 3-sided & 170 & 0.7 & 1.81 &
$0.0917^{+0.0199}_{-0.0137}$ & 500 &
$27.61^{+6.47}_{-5.83}$ & 21.33 & 0.31 \\[1pt]

2025-10-02\_00-40-53 & 3-sided & 170 & 0.4 & 1.82 &
$0.1078^{+0.0196}_{-0.0151}$ & 500 &
$27.01^{+5.40}_{-7.41}$ & 20.30 & 0.37 \\

\hline
\end{tabular}%
}

\renewcommand{\arraystretch}{1}

\end{table*}

\twocolumn
\raggedbottom

\section{Processing the PSF images}
\label{sec:appendix_processing_psf_images}

The PSF images from the C-RED2 were stored as 25-frame cubes, where each frame represents the average of 100 individual short-exposure images. Each frame was pre-processed to mitigate background contributions by subtracting a scaled detector dark and bias frame. The residual background structure, particularly patterns varying along rows and columns, was further removed by estimating and subtracting row-wise and column-wise components, effectively suppressing large-scale fluctuations.

\begin{figure}[H]
\centering
\includegraphics[width=0.5\textwidth]{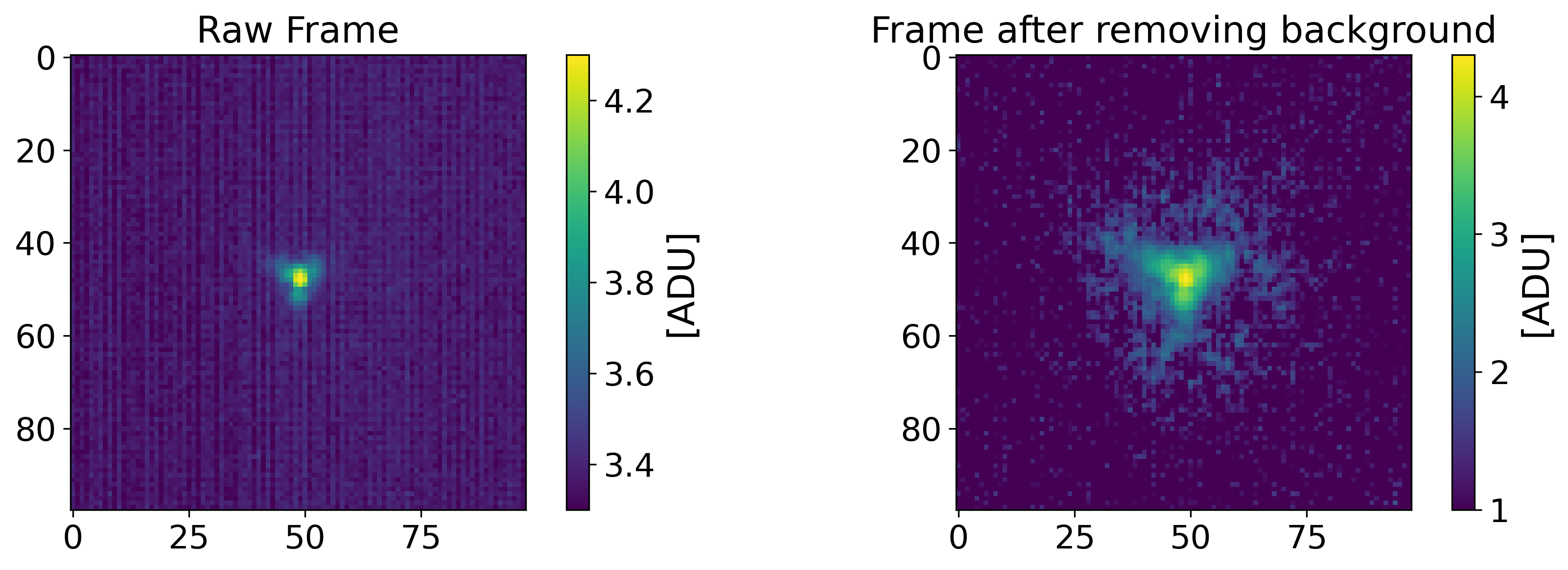}
\caption{Example of a single C-RED2 frame before (left) and after (right) background subtraction. Both images are displayed in a log scale.}
\label{fig:background_subtraction}
\end{figure}

An example of a single raw frame and the corresponding frame after background subtraction is shown in Fig.~\ref{fig:background_subtraction}. The procedure efficiently removes the structured background, yielding a significantly higher-contrast and cleaner PSF.

\section{Tip-tilt jitter}
\label{sec:tip_tilt_jitter}

\begin{figure}[H]
\centering
\includegraphics[width=0.9\columnwidth]{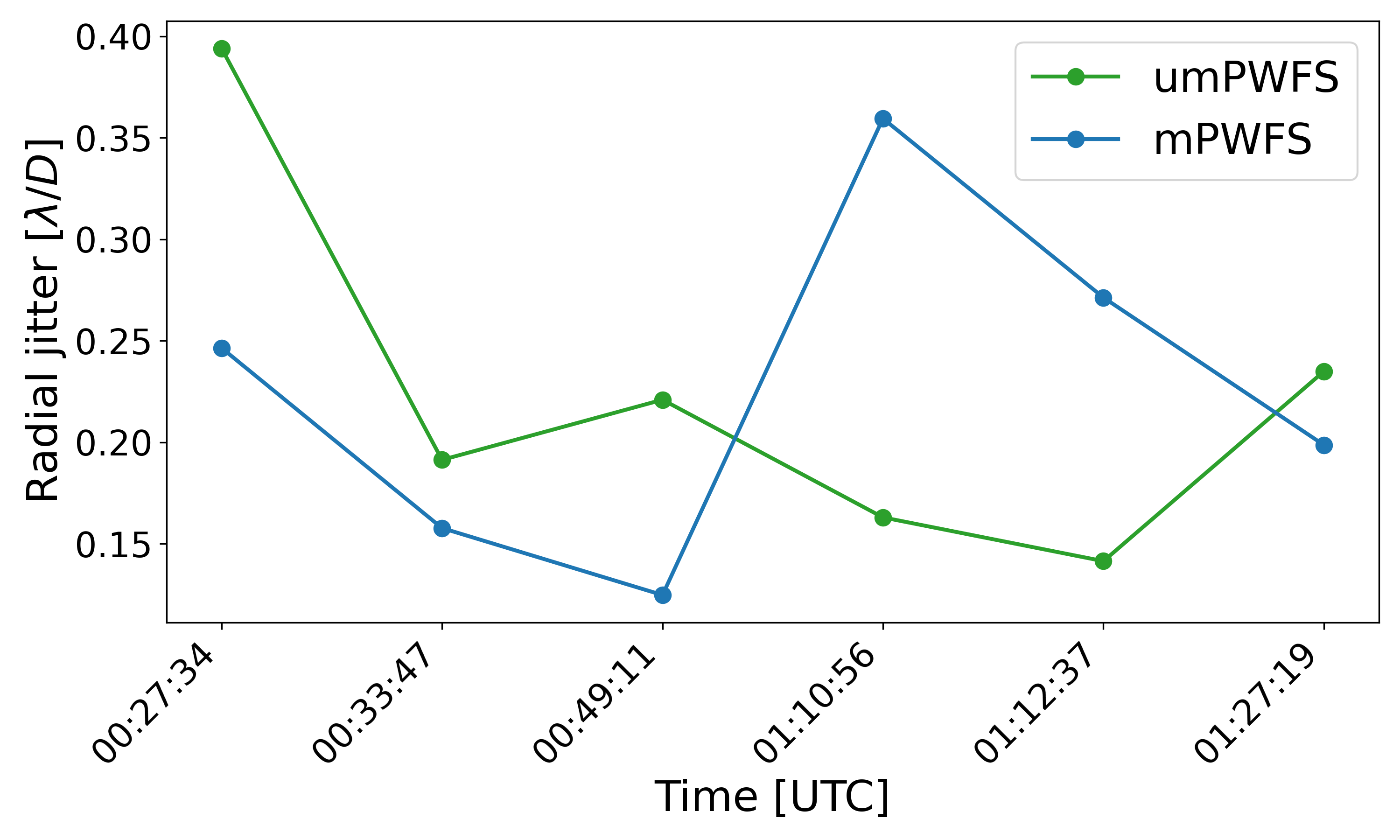}
\caption{Tip-tilt jitter comparison as a function of time for several observations with the IR-umPWFS (green; $1000\,\mathrm{Hz}$) and VIS-mPWFS (blue; $500\,\mathrm{Hz}$). Each time stamp corresponds to the start of the IR-umPWFS observation of a pair; the paired VIS-mPWFS observation was acquired within a few minutes (see Table~\ref{tab:3pwfs_1000_4pwfs_500} for the exact times).}
\label{fig:tip-tilt_jitter_vs_time}
\end{figure}

We computed the tip-tilt jitter of the wavefront sensors in their default configurations: the IR-umPWFS operating at 1000\,Hz and the VIS-mPWFS at 500\,Hz. It is estimated from the temporal displacement of the PSF centroid throughout each observation as a measure of image stability. Because each PSF image corresponds to an average of 100 frames, the measured jitter represents a time-averaged rather than instantaneous tip-tilt motion.
As shown in Fig.~\ref{fig:tip-tilt_jitter_vs_time}, the jitter levels are comparable between the IR-umPWFS and VIS-mPWFS, with no systematic advantage observed for either wavefront sensor. Some individual measurements are affected by transient events. In the first observation pair (00:27:34 UT), the higher jitter measured with the IR-umPWFS is caused by a sudden jump in the PSF position. Similarly, in the fourth pair (01:10:56 UT), the increased jitter of the VIS-mPWFS is due to a brief opening of the loop during the acquisition. 

\section{Optical gains}
\label{sec:optical_gains_appendix}
One of the main limitations of PWFS is its intrinsically nonlinear response, which leads to a spatial-frequency-dependent loss of sensitivity during on-sky operations. This loss can be quantified through the concept of optical gains \citep{Deo2019, Chambouleyron2020}.

The interaction matrix, $\mathcal{D}_{\mathrm{calib}}$, is computed during the calibration process under ideal conditions using a reference image corresponding to a flat wavefront. During on-sky observations, the PWFS operates under different conditions in the presence of residual wavefront error. Due to the nonlinear nature of the WFS, this mismatch leads to a modification of its effective response. To account for these effects, we model the PWFS as a system with a locally linear behavior that depends on the sensing regime. In particular, we assume that the PWFS response in the on-sky regime can be described by an effective interaction matrix, denoted as $\mathcal{D}_{\mathrm{onSky}}$.

For each mode $\phi_i$, the slope of the linear response in the calibration regime, $\delta S_{\mathrm{calib}}(\phi_i)$, is estimated using a push-pull technique:
\begin{equation}
\delta S_{\mathrm{calib}}(\phi_i) = \frac{S_{\mathrm{calib}}(+a \phi_i) - S_{\mathrm{calib}}(-a \phi_i)}{2a},
\end{equation}
where $S_{\mathrm{calib}}$ denotes the PWFS slope defined in Sect.~\ref{sec:simulation_description}, and $a$ is the amplitude of the applied modal perturbation.

Similarly, in the on-sky regime, the response $\delta S_{\mathrm{onSky}}(\phi_i)$ is obtained as:
\begin{equation}
\delta S_{\mathrm{onSky}}(\phi_i) = \frac{S_{\mathrm{onSky}}(+a \phi_i) - S_{\mathrm{onSky}}(-a \phi_i)}{2a},
\end{equation}
where $S_{\mathrm{onSky}}$ represents the PWFS slope measured during closed-loop operation under on-sky conditions.

We define the optical transfer matrix $T_{\mathrm{opt}}$ as the matrix relating the responses in the calibration and on-sky regimes, $\mathcal{D}_{\mathrm{onSky}} = \mathcal{D}_{\mathrm{calib}}\,T_{\mathrm{opt}}$. Following \citet{Deo2019}, we assume that $T_{\mathrm{opt}}$ is diagonal, implying that no modal cross-talk is introduced when transitioning between regimes. Under this assumption, the response of each mode $\phi_i$ is scaled by a scalar factor $G(\phi_i)$, referred to as the modal optical gain.

For each mode $\phi_i$, the optical gain is defined as:
\begin{equation}
G(\phi_i) = \frac{\langle \delta S_{\mathrm{onSky}}(\phi_i), \, \delta S_{\mathrm{calib}}(\phi_i) \rangle}
{\langle \delta S_{\mathrm{calib}}(\phi_i), \, \delta S_{\mathrm{calib}}(\phi_i) \rangle},
\end{equation}

An equivalent matrix formulation is given by:
\begin{equation}
G_{\mathrm{opt}} = \frac{\mathrm{diag}\left( \mathcal{D}_{\mathrm{onSky}}^{\mathrm{T}} \, \mathcal{D}_{\mathrm{calib}} \right)}
{\mathrm{diag}\left( \mathcal{D}_{\mathrm{calib}}^{\mathrm{T}} \, \mathcal{D}_{\mathrm{calib}} \right)},
\end{equation}

where the division is element-wise and $G_{\mathrm{opt}}$ is a vector containing the modal optical gains $G(\phi_i)$ for $i \in [1, N_{\mathrm{modes}}]$.

\begin{figure}
\centering
\includegraphics[width=0.45\textwidth]{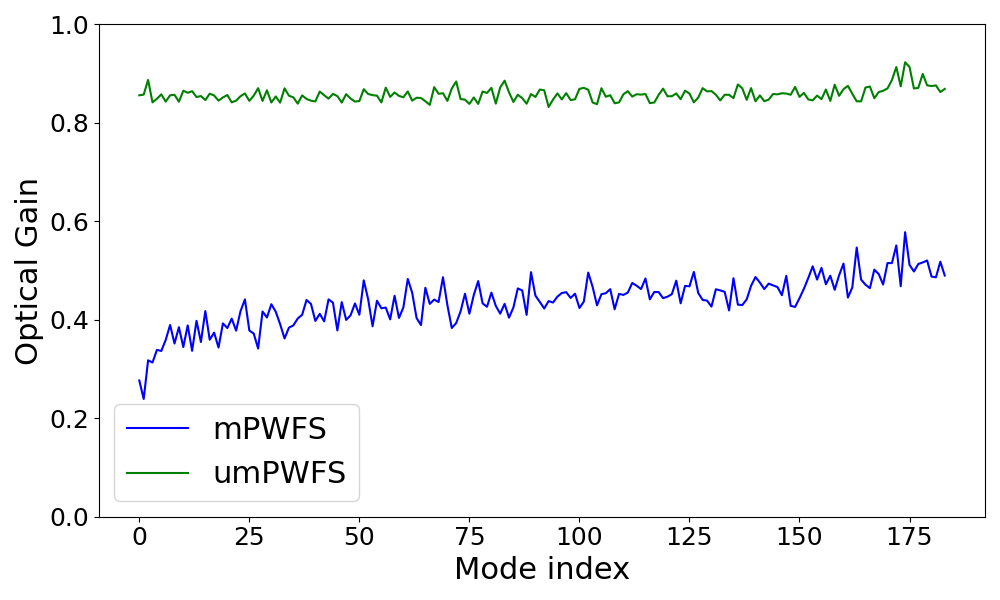}
\caption{Optical gain as a function of mode index for the VIS-mPWFS (500\,Hz; blue) operating at $\lambda_{\mathrm{WFS}} =$ 700\,nm and the IR-umPWFS (1000\,Hz; green) operating at $\lambda_{\mathrm{WFS}} =$ 1400\,nm, using closed-loop residuals at a seeing of $3''$.}
\label{fig:optical_gains}
\end{figure}

Using the end-to-end simulations and configuration described in Sect.~\ref{sec:simulation_description}, we computed the optical gains  $G_{\mathrm{opt}}$  for the IR-umPWFS at 1000\,Hz and the VIS-mPWFS at 500\,Hz (with a modulation radius of $5\,\lambda/D$). The corresponding results are shown in Fig.~\ref{fig:optical_gains}. The optical gains for the IR-umPWFS are nearly flat and high, at $\sim$0.85, while those of the VIS-mPWFS vary between 0.3 and 0.5. A clear drop in optical gain is observed for modes within the modulation radius in the VIS-mPWFS case, whereas the absence of modulation in the IR-umPWFS preserves the sensor sensitivity.

\end{appendix}

\end{document}